\documentclass{tlp}
\usepackage{amssymb}
\usepackage{epsf}

\newtheorem{lemma}{Lemma}
\newtheorem{definition}{Definition}

\newtheorem{example}{Example}
\newtheorem{proposition}{Proposition}
\newtheorem{corollary}{Corollary}

\def\gapa{\hspace{-1pt}}
\def\gapb{\hspace{1pt}}
\def\gap{\hspace{0pt}}

\begin{document}

\title[Efficient Groundness Analysis in Prolog]{Efficient Groundness
Analysis in Prolog} 
\author[Jacob M. Howe and Andy King]{JACOB M. HOWE \\
Department of Computing, City University, London, EC1V~OHB, UK \\
jacob@soi.city.ac.uk\\
\and
ANDY KING \\
Computing Laboratory, University of Kent, CT2 7NF, UK\\
a.m.king@ukc.ac.uk}
\maketitle

\def\land       {\wedge}
\def\lor        {\vee}
\def\liff       {\leftrightarrow}
\def\iff       {\leftrightarrow}
\def\limply     {\rightarrow}
\def\imp     {\rightarrow}
\def\lif        {\leftarrow}
\def\from        {\leftarrow}
\def\lfor       {\forall}
\def\lexist     {\exists}
\def\lfrom      {\leftarrow}
\def\lto        {\rightarrow}

\def\real       {\mathbb{R}}
\def\restrict   {\overline{\exists}}
\def\lup        {\uparrow}
\def\ldown      {\downarrow}
\def\lhull      {\dot{\curlyvee}}
\def\ldotor     {\dot{\vee}}
\def\lproj      {\overline{\lexist}}
\def\stop       {{\hfill $\blacksquare$}}

\def\myshare    {\mbox{\it Sharing\/}}
\def\mymodel    {\mbox{\it model\/}}
\def\myiff      {\mbox{\it iff\/}}
\def\myif       {\mbox{\it if\/}}
\def\mybool     {\mbox{\it Bool\/}}
\def\mypos      {\mbox{\it Pos\/}}
\def\myepos      {\mbox{\it EPos\/}}
\def\mydef      {\mbox{\it Def\/}}
\def\mymon      {\mbox{\it Mon\/}}

\begin{abstract}
Boolean functions can be used to
express the groundness of, and trace grounding dependencies between,
program variables in (constraint) logic programs. In this paper, a
variety of issues pertaining to the efficient Prolog implementation of
groundness analysis are investigated, focusing on the domain of
definite Boolean functions, \mydef.  
The systematic design of the
representation of an abstract domain is discussed in relation to its
impact on the algorithmic complexity of the domain
operations; the most frequently called operations should be the most
lightweight.  This methodology is applied to \mydef,
resulting in a new
representation, together with new algorithms for its domain 
operations utilising previously unexploited properties of \mydef \ -- for
instance, quadratic-time entailment checking.
The iteration strategy driving
the analysis is also discussed and a simple, but very effective, optimisation
of induced magic is described. 
The analysis can be implemented straightforwardly in Prolog
and the use of a non-ground representation results in an efficient,
scalable tool which does not require widening to be invoked,
even on the largest benchmarks. 
An extensive experimental evaluation
is given.
\end{abstract}

\section{Introduction}

Groundness analysis is an important theme of
logic programming and abstract interpretation. 
Groundness analyses identify those program variables 
bound to terms that contain no variables (ground terms).
Groundness information is typically inferred by tracking dependencies
among program variables. These dependencies are
commonly expressed as Boolean functions. For example, 
the function $x \land (y \lif z)$ 
describes a state in which $x$ is definitely ground, and there
exists a grounding dependency such that whenever $z$ becomes ground
then so does $y$. 

Groundness analyses usually track dependencies
using either $\mypos$, the
class of positive Boolean functions
\cite{BS98,BS93,CD95,fec-sei-99,MS93,CLCVH95}, or $\mydef$, the class 
of definite positive functions 
\cite{AMSS98,D91,BHBDJS96,gen-cod-01,how-kin-00}.
$\mypos$ is more expressive than
$\mydef$, but studies have shown that
$\mydef$ analysers can be faster than comparable
$\mypos$ analysers \cite{AMSS98}
and, in practice,
the loss of precision for goal-dependent groundness
analysis is usually small \cite{HACK00}.
This paper is a development of \cite{how-kin-00} and is
an exploration of the representation of Boolean
functions for groundness analysis
and the use of Prolog as
a medium for implementing all the components of a groundness
analyser. 

The rationale for this work was to develop an analyser with
conceptually simple domain operations, with a small and simple (thus easily
maintained) Prolog implementation based on
a meta-interpreter and with performance comparable to that of
BDD-based  analysers.  
Moreover, since Prolog is well suited to symbolic manipulation, it
should provide an appropriate medium for implementing a symbolic analysis, such
as groundness analysis.
Any analysis
that can be quickly prototyped in Prolog is
particularly attractive. The main drawback of this
approach has traditionally been performance. The efficiency
of an analyser can be guaranteed by including a widening (the
controlled exchange of precision for scalability). However,
a successful analyser should
fire widening infrequently to maximise precision.

The efficiency of groundness analysis depends critically on the
way dependencies are represented.  Representation has two aspects: the
theoretical representation (BDDs, Blake Canonical Form, etc.) 
of the Boolean functions and the 
data-structures of the implementation language that are used to support this
representation. 
The theoretical representation determines the complexity of the domain
operations, but the implementation requires the specific data-structures 
used to be amenable to efficient implementation in the
chosen language.  That is, the implementation can push the complexity
into a higher class, or introduce a prohibitive constant factor in the
complexity function.   
This paper considers how a representation should be chosen 
for the intended application (groundness analysis) by balancing the
size of the representation (and its impact) with the complexity of the
abstract operations and the frequency with which these operations are
applied.  The paper also explains how Prolog can be used to implement
a particularly efficient
\mydef-based groundness analysis.
The orthogonal issue of the iteration strategy used to drive the
analysis is also considered.
Specifically, this paper makes the following contributions:
\begin{itemize}
\item  A representation of \mydef \ formulae as non-canonical
conjunctions of clauses is chosen by following a methodology that
advocates: 1) ensuring that the most
commonly called domain operations are the most lightweight; 2) that the
abstractions that arise in practice should be dense; 3)
that, where possible, expensive domain operations should be filtered by
lightweight special cases.
\item A fast Prolog implementation of \mydef-based groundness analysis is
given \linebreak founded on the methodology above, using a compact, factorised 
representation.  
\item Representing Boolean functions as non-ground formulae allows
succinct implementation of domain operations. In particular a
constant-time meet is achieved using difference lists and a quadratic-time
entailment check is built using delay declarations.  
\item A new join algorithm  is presented which does not require
formulae to be preprocessed into a canonical form.
\item The use of entailment checking as a filter for join is described,
as is the use of a filtered projection.
\item Various iteration strategies are systematically compared and 
it is suggested (at least for groundness analysis) that good
performance can be obtained by a surprisingly simple analysis framework.
\item An extensive experimental evaluation of groundness analysis
using a variety of combinations of domains, representations and
iteration strategies is given.
\item As a whole, the work presented in this paper strongly suggests that the
implementor can produce a robust,
fast, precise, scalable analyser for goal-dependent groundness
analysis written in Prolog.  The analyser presented does not require
widening to be applied for any programs in the benchmarks suite. 
\end{itemize}

The rest of the paper is structured as follows:
Section 2 details the necessary preliminaries.
Section 3 reviews the methods used for choosing the representation of
\mydef. It goes on to describe various representations of \mydef \ in
relation to a frequency analysis of the operations; a non-canonical
representation as conjunctions of clauses is detailed.
Section 4 describes a new join algorithm, along with filtering
techniques for join and for projection.
Section 5 discusses a variety of iteration strategies for driving an
analysis.  
Section 6 gives an experimental evaluation of the various combinations
of domain representations and iteration strategy for \mydef \ (and
also for the domains \myepos \ and \mypos). 
Section 7 surveys related work and 
Section 8 concludes.

\section{Preliminaries}

A Boolean function is a function 
$f: \mybool^n \rightarrow \mybool$
where $n \geq 0$.
Let $V$ denote a denumerable universe of variables.
A Boolean function can be
represented by a propositional formula over $X \subseteq V$ where $|X| = n$.
The set of 
propositional formulae over $X$ is denoted by $\mybool_X$.
Throughout this paper, Boolean functions and propositional formulae
are used interchangeably
without worrying about the distinction.
The convention of
identifying a truth assignment with the
set of variables $M$ that it maps to $true$ is also followed.
Specifically, a map ${\psi_X}(M) : \wp(X) \lto \mybool_X$
is introduced defined by:
${\psi_X}(M) = (\land M) \land \neg (\lor \! (X \! \setminus \! M))$.
In addition, the formula $\land Y$ is often abbreviated as $Y$.

\begin{definition} \rm
The (bijective) map 
${\mymodel_X} : \mybool_X \lto \wp(\wp(X))$ is defined by:
${\mymodel_X}(f)$ = 
$\{ M \subseteq X \mid {\psi_X}(M) \models f \}$.
\end{definition}

\begin{example} \rm
If $X = \{x, y\}$, then the function
$\{ \langle true, true \rangle \! \mapsto \! true$,
$\langle true, false \rangle \! \mapsto \! false$,
$\langle false, true \rangle \! \mapsto \! false$,
$\langle false, false \rangle \! \mapsto \! false \}$ can be
represented by the formula $x \land y$. Also,
${\mymodel_X}(x \land y)$ =
$\{ \{ x, y \} \}$
and ${\mymodel_X}(x \lor y)$ =
$\{ \{ x \}, \{ y \}$, $\{ x, y \} \}$.
\end{example}

The focus of this paper is on the use of sub-classes of $\mybool_X$ in
tracing groundness dependencies.  These sub-classes are defined below:

\begin{definition} \rm
A function $f$ is positive iff $X \in \mymodel_X(f)$.
$\mypos_X$ is the set of positive Boolean functions over $X$.
A function $f$ is definite iff 
$M \cap M' \in \mymodel_X(f)$
for all $M, M' \in \mymodel_X(f)$.
$\mydef_X$ is the set of positive functions over $X$ that are 
definite.  
A function $f$ is GE iff $f$ is definite positive and,
where $Y = \cap\mymodel_X(f)$,
for all $M, M' \in \mymodel_X(f)$, $Y\cup (M\setminus M') \in
\mymodel_X(f)$.  
$\myepos_X$ is the set of GE
functions over $X$.
\end{definition}

\noindent
Note that 
$\myepos_X \subseteq \mydef_X \subseteq \mypos_X$. One useful representational property
of $\mydef_X$ is that each
$f \in \mydef_X$ can be described as a conjunction
of definite (propositional) clauses, that is,
$f = \land_{i = 1}^{n} (y_{i} \lif \wedge Y_i)$ \cite{D91}.  Note that the
$y_i$s are not necessarily distinct.  
Finally,
$\mydef$ abbreviates $\mydef_V$. Also notice that $EPos_X = \{ \wedge F \mid F
\subseteq X \cup E_X \}$, where $E_X = \{ x \leftrightarrow y \mid x, y
\in X \}$.

\begin{example} \rm
Suppose $X = \{x, y, z\}$ and consider the following table, which states, 
for some Boolean functions, whether they are in $\myepos_X$,
$\mydef_X$ or $\mypos_X$ and also gives $\mymodel_X$. 

\[
\begin{array}{@{}c@{\;}|@{\;}c@{\;}c@{\;}c@{\;}|@{\;}l@{\;}l@{\;}l@{\;}l@{\;}l@{\;}l@{\;}l@{\;}l@{}}
f                 & \myepos_X & \mydef_X & \mypos_X &
\multicolumn{8}{c}{{\mymodel_X}(f)} \\ \cline{1-12} 
false             & &         &            & \multicolumn{8}{c}{\emptyset} \\
x \land y         & \bullet &\bullet  & \bullet  & 
\{ & & & & \{x,y\}, & & & \{x,y,z\} \}\\
x \lor y          &    &      & \bullet  &
\{ & \{x\}, & \{y\}, & & \{x,y\}, & \{x,z\}, & \{y,z\}, & \{x,y,z\} \}\\
x \lif y          & & \bullet  & \bullet &         
\{ \emptyset, & \{x\}, & & \{z\}, & \{x,y\}, & \{x,z\}, & & \{x,y,z\} \}\\
x \lor (y \lif z) &   &       & \bullet &         
\{ \emptyset, & \{x\}, & \{y\}, & & \{x,y\}, & \{x,z\}, & \{y,z\}, & \{x,y,z\} \}\\ 
true              & \bullet & \bullet  & \bullet  &
\{ \emptyset, & \{x\}, & \{y\}, & \{z\}, & \{x,y\}, & \{x,z\}, & \{y,z\}, & \{x,y,z\} \}
\end{array}
\]
Note, in particular, that $x \lor y$ is not in $\mydef_X$ (since
its set of models is not closed under intersection)
and that  $false$ is neither in $\myepos_X$, nor $\mypos_X$ nor $\mydef_X$. 
\end{example}

\noindent
Defining 
$f_1 \ldotor f_2$ =
$\land \{ f \in \mydef_X \mid f_1 \models f \land f_2 \models f \}$,
the 4-tuple $\langle \mydef_X, \models, \land, \ldotor \rangle$
is a finite lattice \cite{AMSS98}, where
$true$ is the top element and
$\wedge X$ is the bottom element. Existential quantification is
defined by 
Schr\"oder's Elimination Principle, that is, $\lexist x . f = f[x \mapsto true] \ldotor f[x \mapsto false]$.
Note that if $f \in \mydef_X$ then $\lexist x . f \in \mydef_X$ \cite{AMSS98}.

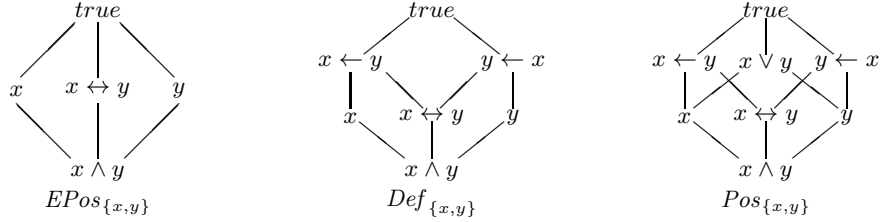
\begin{figure}[t] 
\setlength{\unitlength}{0.12cm}
\begin{center}
\hspace*{-0.5cm}
\begin{tabular}{c@{\qquad\qquad\qquad}c@{\qquad\qquad\qquad}c}

\begin{picture}(19,20)(0,0)
\put(9,-3){\makebox(6,1){$\myepos_{\{ x, y \}}$}}

\put(9,0){\makebox(6,2){$x \land y$}}

\put(9,2.5){\line(-1,1){6}}
\put(12,2.5){\line(0,1){6}}
\put(15,2.5){\line(1,1){6}}

\put(0,9){\makebox(6,2){$x$}}
\put(9,9){\makebox(6,2){$x \liff y$}}
\put(18,9){\makebox(6,2){$y$}}

\put(3.5,11.5){\line(1,1){6}}
\put(12,11.5){\line(0,1){6}}
\put(20.5,11.5){\line(-1,1){6}}

\put(9,18){\makebox(6,2){$true$}}
\end{picture}

&

\begin{picture}(19,20)(0,0)
\put(9,-3){\makebox(6,1){$\mydef_{\{ x, y \}}$}}

\put(9,0){\makebox(6,2){$x \land y$}}

\put(9,2.5){\line(-5,4){5}}
\put(12,2.5){\line(0,1){4}}
\put(15,2.5){\line(5,4){5}}

\put(0,6){\makebox(6,2){$x$}}
\put(9,6){\makebox(6,2){$x \liff y$}}
\put(18,6){\makebox(6,2){$y$}}

\put(3,8){\line(0,1){4}}
\put(11,8){\line(-1,1){4}}
\put(13,8){\line(1,1){4}}
\put(21,8){\line(0,1){4}}

\put(0,12){\makebox(6,2){$x \lif y$}}
\put(18,12){\makebox(6,2){$y \lif x$}}

\put(4.5,14){\line(5,4){5}}
\put(19.5,14){\line(-5,4){5}}

\put(9,18){\makebox(6,2){$true$}}
\end{picture}

& 

\begin{picture}(19,20)(0,0)
\put(9,-3){\makebox(6,1){$\mypos_{\{ x, y \}}$}}

\put(9,0){\makebox(6,2){$x \land y$}}

\put(9,2.5){\line(-5,4){5}}
\put(12,2.5){\line(0,1){4}}
\put(15,2.5){\line(5,4){5}}

\put(0,6){\makebox(6,2){$x$}}
\put(9,6){\makebox(6,2){$x \liff y$}}
\put(18,6){\makebox(6,2){$y$}}

\put(3,8){\line(0,1){4}}
\put(11,8){\line(-1,1){4}}
\put(13,8){\line(1,1){4}}
\put(21,8){\line(0,1){4}}

\put(0,12){\makebox(6,2){$x \lif y$}}
\put(9,12){\makebox(6,2){$x \vee y$}}
\put(18,12){\makebox(6,2){$y \lif x$}}

\put(12,14){\line(0,1){4}}
\put(4,8){\line(5,4){5}}
\put(20,8){\line(-5,4){5}}

\put(4.5,14){\line(5,4){5}}
\put(19.5,14){\line(-5,4){5}}

\put(9,18){\makebox(6,2){$true$}}
\end{picture}

\end{tabular}
\end{center}
\caption{Hasse Diagrams\label{hasse}}
\end{figure}

\begin{example}\label{dotvee-ex} \rm
If $X = \{x,y\}$ then
$x \ldotor (x \liff y)$ =
$\land \{ (x \lif y), true \}$ = $(x \lif y)$,
as can be seen 
in the Hasse diagram for dyadic $\mydef_X$ (Fig.~\ref{hasse}).
Note also that $x \ldotor y$ =
$\land \{ true \}$ = $true \neq (x \lor y)$. 
\end{example}

The set of (free) variables in a syntactic object $o$ is denoted
by $var(o)$. Also,
$\lexist \{ y_1, \ldots, y_n \} . f$ (project out) abbreviates
$\lexist y_1 . \ldots . \lexist y_n . f$ and
$\restrict Y . f$ (project onto) denotes
$\lexist var(f) \setminus Y . f$.  Let $\rho_1, \rho_2$ be fixed renamings
such that 
$X \cap \rho_1(X)$ =
$X \cap \rho_2(X)$ =
$\rho_1(X) \cap \rho_2(X) = \emptyset$. Renamings are bijective and
therefore invertible.

Downward closure, $\ldown$, relates $\mypos$ and $\mydef$
and is useful when tracking sharing with Boolean functions \cite{CSS99}.
It is defined by
$\ldown \!\! f = \mymodel_X^{-1}(\{ \cap S \mid
\emptyset \subset S \subseteq \mymodel_X(f)\})$.
Note
that
$\ldown \!\! f$ has the useful computational property
that $\ldown \!\! f = \land \{ f' \in \mydef_X \! \mid \! f \models f' \}$ if
$f \in \mypos_X$.  That is, $\ldown$ takes a \mypos \ formula to its
best \mydef \ approximation. Finally, for any $f \in \mybool_X$, $coneg(f)$ =
$\mymodel_X^{-1}(\{ X \setminus M \mid M \in \mymodel_X(f) \})$ \cite{CSS99}.


%

The following pieces of logic programming terminology will also be
needed. Let $T$ denote the set of terms constructed from $V$ and a
set of function symbols $F$.  $\Pi$ is a set of predicate
symbols.
An equation $e$ is a pair $(s = t)$
where $s, t \in T$. A substitution is a (total) map $\theta : V
\to T$ such that $\{ v \in V \mid \theta(v) \neq v \}$ is finite.
Let $Sub$ denote the set of substitutions and let $E$ denote a
finite set of equations. Let $\theta(t)$ denote the term obtained
by simultaneously replacing each occurrence of $v$ in $t$ with
$\theta(v)$, and let $\theta(E)$ = $\{ \theta(s) = \theta(t) \mid
(s = t) \in E \}$.

Composition of substitutions induces the (more general than)
relation $\leq$ defined by: $\theta \leq \psi$ if there exists
$\delta \in Sub$ such that $\psi = \delta \circ \theta$. More
general than lifts to terms by $s \leq t$ iff there exists $\delta
\in Sub$ such that $\delta(s) = t$. The set of unifiers of $E$,
$unify(E)$, is defined by: $unify(E)$ = $\{ \theta \in Sub \mid
\lfor (s = t) \in E . \theta(s) = \theta(t) \}$ and the set of
most general unifiers, $mgu(E)$, is defined by: $mgu(E) = \{
\theta \in unify(E) \mid \lfor \psi \in unify(E) . \theta \leq
\psi \}$. Finally, the set of generalisations of two terms is
defined by: $gen(t_1, t_2) = \{ t \in T |  t \leq t_1 \wedge t
\leq t_2\}$ and the set of most specific generalisations is
defined by: $msg(t_1, t_2) = \{ t \in gen(t_1, t_2) | \forall s
\in gen(t_1, t_2). s \leq t \}$.

\section{Choosing a Representation for \mydef}

\subsection{Review of Design Methods}

The efficiency of an analyser depends critically on the algorithmic
complexities of
its abstract domain operations.  
These in turn are determined by the
representation of the abstract domain.
The representation also 
determines the 
size of the inputs to the domain operations, as well as impacting on
memory usage.
Because of this, the choice of
representation is fundamental to the efficiency of an analyser and
is therefore of great importance.  
The remainder of this subsection reviews
three factors which should help the
implementor arrive at a suitable representation and suggest where
domain operations might be refined.

\subsubsection{Frequency Analysis of the Domain Operations}

There are typically many degrees of freedom in designing
an analyser, even for a given domain.
Furthermore, work can often be shifted from one abstract
operation into another.  
For example, Boolean formulae can be represented in either conjunctive
normal form (CNF) or disjunctive normal form (DNF).  In CNF,
conjunction is constant-time and disjunction is quadratic-time,
whereas in DNF, conjunction is quadratic-time and disjunction is
constant-time.  
Ideally, an analysis should be designed so
that the most frequently used operations have low complexity
and are therefore fast. This motivates the following approach:
\begin{enumerate}
\item Prototype an analyser for the given domain.
\item Instrument the analyser to count the number of times each
domain operation is invoked.
\item Generate these counts for a number of programs (the bigger the better). 
\item Choose a representation which gives a good match between
the frequency and the complexity of the domain operations.
\end{enumerate}

\noindent Because the frequency analysis is solely concerned with generated
instruction counts, the efficiency of the prototype analyser is not a
significant issue.  The objective is to choose a representation for
which the most frequently occurring operations are also the fastest.
However, this criterion needs to be balanced with others, such as the
density of the representation.

\subsubsection{Density of the Domain Representation}

The complexity of the domain operations is a function of the size of
their inputs.
Large inputs nullify the value of good complexities, hence a balance
between size of representation and complexity of domain operations
is needed. The following factors impact on this relationship:
\begin{enumerate}
\item The abstractions which typically arise should be
represented compactly.
\item A factorised representation with an expressive, high density, low
maintenance component is attractive.
\item Maintaining the representation (for example, as a canonical form)
should not come with a high overhead.
\item The representation should fit with machinery available in the
implementation language.
\end{enumerate}
\noindent A domain is said to be factorised if its information is
represented as a product of subdomains.  It may not always be possible
to fulfill all these requirements.
Moreover, these factors needs to balanced with others, such as their
impact on the complexities of frequently called domain operations.

\subsubsection{Filtering the Domain Operations}

In many analyses it is inevitable that some domain operations will
have high complexity.  However, it is sometimes possible to reduce the
impact of this by filtering the operation, as follows:
\begin{enumerate}
\item For a high complexity domain operation identify special cases
where the operation can be
calculated using a lower complexity algorithm.
\item Instrument the analyser to quantify how often the lower
complexity algorithm can be applied.
\item If it appears that the special case occurs frequently, then
implement the special case and measure the impact on performance.
\end{enumerate}
\noindent The bottom line is that the cost of detecting the special
case should not outweigh the benefit of applying the specialised
domain algorithm.

%

\subsection{Frequency Analysis for \mydef}

In order to balance the frequency of abstract operations against their cost,
an existing \mydef \ analyser was instrumented to count the number of calls
to the various abstract operations.   The analyser used for this is based 
on Armstrong and Schachte's BDD-based domain operations for \mypos \
and \myshare \ 
\cite{AMSS98, sch-phd}.  Using the domain operations provided for
these domains, a \mydef \ analyser can easily be derived.
This analyser
is coded in Prolog as a simple
meta-interpreter that uses induced magic-sets
\cite{Codish:fastmagic} and eager evaluation \cite{Wund:lobstre95}
to perform goal-dependent bottom-up evaluation and call the C implemented
domain operations.

Induced magic is a refinement of the 
magic set transformation,
avoiding much of the re-computation that 
arises because of the repetition of literals in the
bodies of magicked clauses \cite{Codish:fastmagic}.
Eager evaluation \cite{Wund:lobstre95} is
a fixpoint iteration strategy which proceeds as follows:
whenever an atom is updated with a new (weaker)
abstraction, a recursive procedure is invoked to ensure that every
clause that has that atom in its body is re-evaluated. 
An advantage of induced magic is that it can be coded
straightforwardly in Prolog.


Table 1 gives a breakdown of the relative frequency (in percentages) of the
calls to each abstract operation in 
the BDD-based $\mydef$ analysis of eight large programs.
Meet, join, equiv, project and rename are the obvious Boolean
operations. 
Join (diff) is those
calls to a join $f_1 \ldotor f_2$ where $f_1 \ldotor f_2 \neq f_1$
and $f_1 \ldotor f_2 \neq f_2$ (this will be useful in section \ref{filt}).
Total details the total number of calls to these domain operations.

\setlength{\tabcolsep}{2pt}
\begin{table}[h!]\label{freq-bdd}
\begin{tabular}{@{}r|@{}r|@{}r|@{}r|@{}r|@{}r|@{}r|@{}r|@{}r@{}}
file            & rubik & chat\_parser & sim\_v5-2 & 
peval & aircraft & essln & chat\_80 & aqua\_c \\ \cline{1-9}
meet        & 30.9 & 31.6  & 35.9 & 32.5 & 28.5  & 42.7  & 34.0  & 34.2 \\
join        & 10.4 & 10.4  &  8.8 &  9.7 & 11.1  &  8.4  & 10.2  & 10.5 \\
join (diff) &  1.1 &  1.7  &  0.0 &  2.9 &  0.1  &  0.9  &  1.5  &  1.6 \\
equiv       & 10.4 & 10.4  &  8.8 &  9.7 & 11.1  &  8.4  & 10.2  & 10.5 \\
project     & 12.6 & 12.5  & 13.0 & 12.5 & 13.0  & 10.5  & 12.1  & 11.7 \\
rename      & 34.7 & 33.4  & 33.6 & 32.8 & 36.2  & 29.2  & 32.0  & 31.6 \\
\cline{1-9}
total       & 14336 & 14124 & 5943 & 6275 & 24758 & 19051 & 45444 & 280485
\end{tabular}
\caption{Frequency Analysis: BDD-based \mydef Analyser (Figures in \%)}
\end{table}

Observe that meet and rename are 
called most frequently.
Join, equiv and project are called with a similar frequency, but less
frequently than meet and rename.  
Note
that it is rare for 
a join to differ from both its arguments. 
Join is always followed by
an equivalence and this explains why the join and equiv rows coincide.
Since meet and rename are the most frequently called operations,
ideally they should be the most lightweight.  
As join, equiv and project are called less frequently,
a higher algorithmic complexity is more acceptable for these
operations.  

\subsection{Representations of Def}

This section reviews a number of representations of \mydef \ in terms
of the algorithmic complexity of the domain operations.  
The representations considered are reduced ordered binary decision
diagrams, dual Blake canonical form (specialised for
$\mydef$ \cite{AMSS98}) and a non-canonical definite propositional clause
representation.   
\begin{description}
\item[\it{ROBDD}]A reduced ordered binary decision diagram (ROBDD) is a rooted,
directed acyclic graph. 
Terminal nodes are labelled 0 or 1 and non-terminal nodes are labelled
by a variable and have edges directed towards two child nodes.
\linebreak ROBDDs
have the additional properties that: 1) 
each path from the root to a node respects a given ordering on the
variables, 2) a variable cannot occur multiply in a path, 3)  no
subBDD occurs multiply.  
ROBDDs give a unique
representation for every Boolean function (up to
variable ordering).
\item[\it{DBCF}]Dual Blake Canonical Form (DBCF) 
represents \mydef \ 
functions as conjunctions of definite (propositional) clauses
\cite{AMSS98, D91, BHBDJS96}
maintained in a canonical (orthogonal) form that
makes explicit transitive variable dependencies and uses a reduced monotonic
body form.  For example, the
function $(x \lif y) \land (y \lif z)$ is
represented as $(x \lif (y \lor z)) \land (y \lif z)$. Again, DBCF
gives a unique representation for every \mydef \  function (up to
variable ordering).
\item[\it{Non-canonical}]The non-canonical clausal representation expresses \mydef \ functions
as conjunctions of propositional clauses, but does not maintain a
canonical form.  This does not give a unique representation.
\end{description}

Table 2 details the complexities of the domain operations for \mydef \
for the three representations.  Notice that the complexities are in
terms of the size of the representations and that these are all
potentially exponential in the number of variables.  Also, observe
that since DBCF maintains transitive dependencies, whereas the
non-canonical representation does not, the DBCF of a \mydef \
function has the potential to be considerably larger than the
non-canonical representation.  As ROBDDs are represented in a
fundamentally different way, their size cannot be directly compared 
with clausal representations.

\begin{table}
\begin{tabular}{|l||c|c|c|c|c|}
\cline{1-6}
Representation & meet & join & equiv & rename & project \\
\cline{1-6}
ROBDD & $O(N^2)$ & $O(N^2)$ & $O(1)$ & $O(N^2)$ & $O(N^2)$ \\
DBCF & $O(N^4)$ & $O(2^{2N})$ & $O(N)$ & $O(N)$ & $O(N)$ \\
Non-canonical & $O(1)$ & $O(2^{2N})$ & $O(N^2)$ & $O(N)$ & $O(2^N)$ \\
\cline{1-6}
\end{tabular}
\caption{Complexity of \mydef \  Operations for Various
Representations (where $N$ is the size of the representation -- number
of nodes/variable occurrences).}
\end{table}

Both ROBDDs and DBCF are maintained in a canonical form.  Canonical forms
reduce the cost of operations such as equivalence checking and
projection by factoring out search.  However, canonical forms need to be
maintained and this maintenance has an associated cost in meet and
join.  That is, ROBDDs and DBCF buy low complexity equivalence
checking and projection at the cost of higher complexity meet and
join.

As discussed in the previously, the lowest cost operations
should be those that are most frequently called.  Table 1 shows that
for \mydef \ based groundness analysis, meet and renaming are called
significantly more often than the other operations.  Hence these
should be the most lightweight.  
This suggests that the non-canonical
representation is better suited to \mydef-based goal-dependent
groundness analysis 
than ROBDDs and DBCF.   
The following sections will detail the non-canonical representation.

\subsection{GEP Representation}

This section outlines how the non-canonical representation is used in
an analysis for call and answer patterns. 
Implementing call and answer patterns with a non-ground representation
enables the non-canonical representation to be factorised at little
overhead.

A call (or answer) pattern is a pair $\langle a, f \rangle$ where
$a$ is an atom and $f \in \mydef$. Normally
the arguments of $a$ are distinct variables. The
formula $f$ is a conjunction (list) of propositional
clauses.
In a non-ground representation 
the arguments of $a$ can be instantiated and
aliased to express simple dependency information 
\cite{HACK00}. For example, if $a = p(x_1, ..., x_5)$,
then the atom $p(x_1, true, x_1, x_4, true)$
represents $a$ coupled with the formula
$(x_1 \liff x_3) \land x_2 \land x_5$. This
enables the abstraction
$\langle p(x_1, ..., x_5), (x_1 \liff x_3) \land x_2 \land x_5 \land
(x_4 \imp x_1) \rangle$ to
be collapsed to $\langle p(x_1, true, x_1, x_4, true), x_4 \imp x_1 \rangle$. 
This encoding leads to a more
compact representation and is similar to the GER factorisation of
ROBDDs proposed by Bagnara and Schachte \cite{BS98}. 
The representation of call and answer patterns described above
is called GEP (groundness, 
equivalences and
propositional clauses) where the atom captures the first two
properties and 
the formula the latter.  

Formally, let $GEP = \{\langle p(t_1, ..., t_n), f\rangle \mid
p\in \Pi, t_i\in V \cup \{true\}, f\in (Def\setminus GE)\cup\{true\}\}$.
Define $\models$ by $\langle
p(\vec{a}_1), f_1\rangle
 \models \langle p(\vec{a}_2), f_2\rangle$ iff $\restrict
\vec{x}.((\vec{a}_1 \leftrightarrow
\vec{x}) \wedge f_1) \models
\restrict\vec{x}.((\vec{a}_2 \leftrightarrow
\vec{x}) \wedge f_2)$ and $var(\vec{x}) \cap (var(\vec{a}_1) \cup
var(\vec{a}_2) \cup var(f_1) \cup var(f_2)) = \emptyset$.  
Then $\langle GEP, \models \rangle$ is a preorder.
The
preorder induces the equivalence
relation $\equiv$ defined by $\langle
p(\vec{a}_1), f_1\rangle
 \equiv \langle p(\vec{a}_2), f_2\rangle$ iff $\langle
p(\vec{a}_1), f_1\rangle
 \models \langle p(\vec{a}_2), f_2\rangle$ and $\langle
p(\vec{a}_2), f_2\rangle
 \models \langle p(\vec{a}_1), f_1\rangle$.  
Let $GEP_{\equiv}$ denotes GEP quotiented by the equivalence.
Define $\wedge:
GEP_{\equiv} \times GEP_{\equiv} \to GEP_{\equiv}$ by $[\langle a_1, f_1
\rangle]_{\equiv} \wedge [\langle a_2, f_2 \rangle]_{\equiv} = [\langle
\theta(a_1),
\theta(f_1) \wedge \theta(f_2)\rangle]_{\equiv}$, where $\theta \in
mgu(a_1, a_2)$. 
Then $\langle GEP_{\equiv},
\models, \wedge\rangle$ is a finite lattice.   

The meet of the pairs
$\langle p(\vec{a}_1), f_1 \rangle$ and
$\langle p(\vec{a}_2), f_2 \rangle$ can be computed by
unifying $a_1$ and $a_2$ and concatenating $f_1$ and $f_2$.  The
unification is nearly linear in the arity of $p$ (using
rational tree unification \cite{jaf-84}) and
concatenation is constant-time (using difference lists).  Since the
arguments $\vec{a}_1$ and $\vec{a}_2$ 
are necessarily distinct, the
analyser would unify $\vec{a}_1$ and
$\vec{a}_2$ even in a non-factorised
representation, hence no extra overhead is incurred.
The objects that require renaming
are formulae and call (answer) pattern
GEP pairs. 
If a dynamic database is used to
store the pairs \cite{HWD92}, then renaming is automatically applied
each time a pair is looked-up in the database. Formulae
can be renamed with a single call to the Prolog builtin copy\_term.
Renaming is therefore linear.

The GEP factorisation defined above is true, that is, all the GE
dependencies are factored into the atom.  An alternative definition
would be $GEP = \{\langle p(t_1, ..., t_n), f\rangle \mid
p\in \Pi, t_i\in V \cup \{true\}, f\in \mydef\}$.  Here the
factorisation is not necessarily true, 
in the sense that GE dependencies may 
exists in the P component, e.g. $\langle p(x, x, true), true\rangle$
may also be correctly expressed as $\langle p(u, v, w), (u\iff v)
\wedge w\rangle$.  A non-true factorisation may be advantageous when
it comes to implementing the domain and from henceforth GEP will refer
to the non-true factorisation version unless stated otherwise.
The P component may contain redundant (indeed, repeated) clauses and
these may impact adversely on performance.   In
order to avert unconstrained growth of P, a redundancy removal step may
be applied to P at a 
convenient point (via entailment checking).
Since the non-canonical formulae do not need to be
maintained in a  canonical form and since the factorisation is not
necessarily true, the representation is flexible in that it can be
maintained on demand, that is, the implementor can choose to move
dependencies from P into GE at exactly those points in the analysis
where true factorisation gives a performance benefit.

\section{Filtering and Algorithms}\label{filt}

The non-canonical representation has high cost join and projection
algorithms.  Therefore it is sensible to focus on improving the
efficiency of these operations.  This is accomplished through
filtering following the strategy described in section 3.1.  This section
presents a new approach to calculating join and describes the use of
entailment checking as a filter in the join algorithm.  It also
describes a filtering method for projection.

\subsection{Join}

This section describes a new approach to calculating join, inspired by
a convex hull algorithm for polyhedra used in disjunctive constraint
solving \cite{BB93}.
The new join algorithm is first described for formulae and is then
lifted to the GEP representation.

\subsubsection{Join for Formulae}

Calculating join in \mydef \ is not straightforward.  It is not
enough to take the join each possible pair of clauses and conjoin
them -- transitive dependencies also need to be taken into
account. This is illustrated by the following example 
(adapted from \cite{AMSS98}).

\newpage
\begin{example} \rm
Put 
$f_1 = (x \lif u)\wedge (u \lif y)$ 
and
$f_2 = (x \lif v) \wedge (v \lif y)$.
Then $f_1\ldotor f_2 = 
(x \lif (u \land v)) \land (x \lif y)$.
The clause $(x\from (u \wedge v))$ comes from $(x\from
u)\ldotor(x\from v)$, but the clause $x\from y$ is not the result of
the join of any pair of clauses in $f_1, f_2$.  It arises since
$f_1\models x \from y$ and $f_2\models x \from y$, that is, from
clauses which appear in transitive closure.
\end{example}
\noindent
One way in which to address the problem of ensuring that the
transitive dependencies are captured is to make the explicit in the
representation (this
idea is captured in the orthogonal form requirement of
\cite{AMSS98}). However, this leads to redundancy in the formula which
ideally should be avoided. 

%
%

It is insightful to consider $\ldotor$ as an operation
on the models of $f_1$ and $f_2$. Since both $\mymodel_X(f_i)$
are closed under intersection, $\ldotor$
essentially needs to extend $\mymodel_X(f_1) \cup \mymodel_X(f_2)$
with new models $M_1 \cap M_2$ where $M_i \in \mymodel_X(f_i)$ to
compute $f_1 \ldotor f_2$.
The following definition expresses this observation and
leads to a new way of computing $\ldotor$ in terms of
meet, renaming and projection, that does not
require formulae to be first put into orthogonal form.

\begin{definition} \rm The map $\lhull : 
{\mybool_X}^2 \lto {\mybool_X}$ is defined by:
$f_1 \lhull f_2 = 
\restrict Y . f_1 \! \curlyvee \! f_2$ where
$Y \! = \! var(f_1) \cup var(f_2)$ and
$f_1 \! \curlyvee \! f_2 \! = \! {\rho_1}(f_1) \land
{\rho_2}(f_2) \land
\land_{y \in Y} y \liff (\rho_1(y) \land \rho_2(y))$.
\end{definition}

\noindent
The following example illustrates the $\lhull$ operator.

\begin{example}
Let $f_1 = (x\from u) \wedge (u\from y)$, $f_2 = (x\from v) \wedge
(v\from y)$.  Then $Y=\{u, v, x, y\}$.  The following substitutions
rename the functions apart, $\rho_1 = \{u \mapsto u', v\mapsto v', x
\mapsto x', y\mapsto y'\}$, $\rho_2 = \{u \mapsto u'', v\mapsto v'', x
\mapsto x'', y\mapsto y''\}$.  Using Definition 3, $f_1\curlyvee f_2 =
(x'\from u') \wedge (u'\from y') \wedge (x''\from v)'' \wedge
(v''\from y'') \wedge u\iff(u'\wedge u'')\wedge v\iff(v'\wedge
v'')\wedge x\iff(x'\wedge x'')\wedge y\iff(y'\wedge y'')$.  Projection
onto $Y$ gives $f_1\lhull f_2 = \lproj\{u, v, x, y\}.f_1\curlyvee f_2
= (x\from (u\wedge v))\wedge (x\from y)$.
\end{example}

\noindent
Note that $\lhull$ operates on ${\mybool_X}$ rather
than $\mydef_X$. This is required for the downward
closure operator in section 5.3. Lemma~\ref{lemma-model} expresses
a key relationship between $\lhull$ and the models of
$f_1$ and $f_2$.

\begin{lemma}\label{lemma-model} \rm
Let $f_1, f_2 \in \mybool_X$.
$M \in \mymodel_X(f_1 \lhull f_2)$ if and only if
there exists 
$M_1 \in \mymodel_X(f_1)$ and $M_2 \in \mymodel_X(f_2)$ such that $M =
M_1 \cap M_2$. 
\end{lemma}

\begin{proof}
Put $X' = X \cup \rho_1(X) \cup \rho_2(X)$.  
Let $M \in \mymodel_X(f_1 \lhull f_2)$.  There exists $M \subseteq M'
\subseteq X'$
such that
$M' \in \mymodel_{X'}(f_1 \curlyvee f_2)$.
Let $M_i = \rho_i^{-1}(M' \cap \rho_i(Y))$, for $i\in \{1, 2\}$.  Thus
$M_i\in model_X(f_i)$ for $i\in\{1, 2\}$.
Observe that $M \subseteq M_1 \cap M_2$ since
$f_1\curlyvee f_2 \models y \imp (\rho_1(y) \land \rho_2(y))$. Also
observe that $M_1 \cap M_2 \subseteq M$ 
since $f_1\curlyvee f_2 \models (\rho_1(y) \land \rho_2(y))\imp y$. Thus
$M = M_1 \cap M_2$, as required.

Let $M_i \in \mymodel_X(f_i)$ for $i\in\{1, 2\}$ and put 
$M = M_1 \cap M_2$ and
$M' = M \cup \rho_1(M_1) \cup \rho_1(M_2)$.
Observe $M' \in \mymodel_{X'}(
f_1 \curlyvee f_2)$ so that
$M \in \mymodel_{X}(f_1 \lhull f_2)$. 
\end{proof}

\noindent
From lemma~\ref{lemma-model} flows the following
corollary and also the useful result that $\lhull$ is monotonic.

\begin{corollary} \rm
Let $f \in \mypos_X$. Then $f = f \lhull f$ if and only if $f \in \mydef_X$. 
\end{corollary}

%

\begin{lemma}\label{lemma-mono} \rm
$\lhull$ is monotonic, that is, 
$f_1 \lhull f_2 \models f'_1 \lhull f'_2$ whenever 
$f_1 \models f'_1$ and $f_2 \models f'_2$.
\end{lemma} 

\begin{proof}
Let $M \in \mymodel_X(f_1 \lhull f_2)$.
By lemma~\ref{lemma-model}, there exist
$M_i \in \mymodel_X(f_i)$ such that $M = M_1 \cap M_2$.
Since
$f_i \models f'_i$, $M_i \in \mymodel_X(f'_i)$ and hence,
by lemma~\ref{lemma-model}, $M \in \mymodel_X(f'_1 \lhull f'_2)$.
\end{proof}

\noindent
The following proposition states that $\lhull$ coincides with
$\ldotor$ on $\mydef_X$. This gives a simple algorithm for
calculating $\ldotor$ that does not depend on the
representation of a formula.

\begin{proposition}\label{prop-equiv} \rm
Let $f_1, f_2 \in \mydef_X$. Then
$f_1 \lhull f_2 = f_1 \ldotor f_2$.
\end{proposition}

\begin{proof}
Since $X \models f_2$ it follows by monotonicity that $f_1 =
f_1 \lhull X \models f_1 \lhull f_2$ and similarly
$f_2 \models f_1 \lhull f_2$. 
Hence $f_1 \ldotor f_2 \models f_1 \lhull f_2$ by
the definition of $\ldotor$.

Now let $M \in \mymodel_X(f_1 \lhull f_2)$. 
By lemma~\ref{lemma-model}, there exists
$M_i \in \mymodel_X(f_i)$ such that
$M = M_1 \cap M_2 \in \mymodel_X(f_1 \ldotor f_2)$.
Hence $f_1 \lhull f_2 \models f_1 \ldotor f_2$. 
\end{proof}

\subsubsection{Join for GEP}

Join, $\vee:
GEP_{\equiv} \times GEP_{\equiv} \to GEP_{\equiv}$,  in the GEP representation can be defined in terms of
$\wedge$ and $\models$ in the usual way, i. e.
$$ 
[\langle a_1, f_1
\rangle]_{\equiv} \vee [\langle a_2, f_2 \rangle]_{\equiv} = 
\bigwedge \left\{ [\langle a, f\rangle]_{\equiv} \in GEP_{\equiv} \ \left| 
\begin{array}{l}
{[\langle a_1, f_1\rangle]}_{\equiv} \models [\langle a,
f\rangle]_{\equiv},  \\
{[\langle a_2, f_2\rangle]}_{\equiv}\models
[\langle a, f\rangle]_{\equiv} 
\end{array} \right. \right\}$$ 
In practice quotienting manifests itself through the 
dynamic database.  Each time a pattern is read from the database it is
renamed.  Join is lifted to quotients by  
reformulated GEP pairs as follows:  $\langle p(\vec{a}_1), f_1 \rangle$ 
becomes $\langle p(\vec{a}), (\vec{a} \iff \vec{a}_1) \wedge f_1 \rangle$ 
where $p(\vec{a}) = msg(p(\vec{a}_1), p(\vec{a}_2))$.
$p(\vec{a})$ is computed using
Plotkin's anti-unification algorithm 
in $O(N\log(N))$ time, where $N$ is the arity of $p$ \cite{P70}.
The following lemma formalises
this lifting of the join algorithm to the 
GEP representation.

\begin{lemma}\label{join-lem}
$[\langle p(\vec{t}_1), f_1\rangle]_{\equiv} \vee [\langle p(\vec{t}_2),
f_2\rangle]_{\equiv}  = [\langle p(\vec{t}), 
(f_1 \wedge (\vec{t}_1
\leftrightarrow \vec{t})) \lhull (f_2 \wedge (\vec{t}_2
\leftrightarrow \vec{t})) 
\rangle]_{\equiv}$, where $\vec{t}\in msg(\vec{t}_1, \vec{t}_2)$.
\end{lemma}

\newpage
\begin{proof}
The first equality holds 
by the definition of $\equiv$ in $GEP_{\equiv}$, the second 
by the definition of join in $GEP_{\equiv}$, the third by the
definition of $\models$ in $GEP_{\equiv}$, the fourth by the
definition of $\wedge$ in $GEP_{\equiv}$,  and the last by
Proposition \ref{prop-equiv}.
$$\begin{array}{lll}
\multicolumn{3}{l}{[\langle p(\vec{t}_1), f_1\rangle]_{\equiv} \vee [\langle p(\vec{t}_2),
f_2\rangle]_{\equiv}} \\
\quad & 
= & [\langle p(\vec{t}), (\vec{t}_1 \iff \vec{t}) \wedge
f_1\rangle]_{\equiv} \vee [\langle p(\vec{t}), (\vec{t}_2 \iff \vec{t}) \wedge
f_2\rangle]_{\equiv} \\
\quad & = &
\wedge\left\{[\langle p(\vec{t}'), f' \rangle]_{\equiv}
\left| \begin{array}{l} {[\langle p(\vec{t}), (\vec{t}_1 \iff \vec{t})\wedge f_1\rangle]}_{\equiv}
\models {[\langle p(\vec{t}'), f'\rangle]}_{\equiv}, \\ {[\langle p(\vec{t}), (\vec{t}_2 \iff \vec{t})\wedge f_2\rangle]}_{\equiv}
\models {[\langle p(\vec{t}'), f'\rangle]}_{\equiv} \end{array}\right.\right\} \\
\quad & = & \wedge\left\{[\langle p(\vec{t}), f' \rangle]_{\equiv}
\left| \begin{array}{l} {[\langle p(\vec{t}), (\vec{t}_1 \iff \vec{t})\wedge f_1\rangle]}_{\equiv}
\models {[\langle p(\vec{t}), f'\rangle]}_{\equiv}, \\ {[\langle p(\vec{t}), (\vec{t}_2 \iff \vec{t})\wedge f_2\rangle]}_{\equiv}
\models {[\langle p(\vec{t}), f'\rangle]}_{\equiv} \end{array}\right.\right\} \\
\quad & = & [\langle p(\vec{t}), \wedge\{f'\in \mydef \mid (\vec{t}_1\iff \vec{t})
\wedge f_1 \models f', (\vec{t}_2 \iff \vec{t}) \wedge f_2 \models
f'\}\rangle]_{\equiv} \\
\quad & = & [\langle p(\vec{t}), (f_1 \wedge (\vec{t}_1
\leftrightarrow \vec{t})) \lhull (f_2 \wedge (\vec{t}_2
\leftrightarrow \vec{t})) \rangle]_{\equiv}
\end{array}
$$
\end{proof}

\subsection{Filtering Join using Entailment Checking}

In section 3.3 it was observed that some high complexity domain operations
have special cases where the operation can be calculated using a lower
complexity algorithm.  Join for \mydef \ in the non-canonical
GEP representation is one such
operation. Specifically, $\ldotor$ is exponential (see Table 2), however,
if $f_1\models f_2$, then $f_1 \ldotor f_2 = f_2$.  Entailment checking is
quadratic in the number of variable occurrences (using a forward
chaining algorithm), hence by using this test, join can be refined.
Table 1 shows that the majority of calls to join will be caught by the
cheaper entailment checking case.  
The following proposition explains how this filtering is lifted to the
GEP representation.  Observe that this proposition has three cases.
The third case is when the entailment check fails.  The first case is
when entailment checking reduces to a lightweight match on the GE
component followed by an entailment check on the P component.   The
second case is more expensive, requiring a most specific
generalisation to be computed as well as an entailment check on more
complicated formulae.  In the context of the analyser, the pair 
$[\langle p(\vec{t}_2), f_2\rangle]_{\equiv}$ corresponds to an
abstraction in the database and these abstractions have the property
that the variables in the P component are contained in those of the GE
component.  This is not necessarily the case for $[\langle
p(\vec{t}_1), f_1\rangle]_{\equiv}$, since in the induced magic
framework $f_1$ represents the
state of the variables  of the clause to the left of the call to 
$p(\vec{t}_1)$. Variable disjointness follows since renaming
automatically occurs every time a fact is read from the dynamic database.
   
\begin{proposition}
Suppose $var(f_2)\subseteq var(p(\vec{t}_2))$ and $var(\langle
p(\vec{t}_1), f_1\rangle) \cap var(\langle p(\vec{t}_2),
f_2\rangle) = \emptyset$.  Then, \\
$[\langle p(\vec{t}_1), f_1\rangle]_{\equiv} \vee [\langle p(\vec{t}_2),
f_2\rangle]_{\equiv}   $
$$=\left\{\begin{array}{lll}
{[\langle p(\vec{t}_2), f_2\rangle]}_{\equiv} & \mbox{if} &
\theta\in mgu(p(\vec{t}_1), p(\vec{t}_2)), \\
& & p(\vec{t}_1) = \theta(p(\vec{t}_2)), \\
& & \theta(f_1) \models \theta(f_2) \\
{[\langle p(\vec{t}_2), f_2\rangle]}_{\equiv} & \mbox{if} & p(\vec{t})
\in msg(p(\vec{t}_1), p(\vec{t}_2)), \\
& & f_1 \wedge (\vec{t}_1 \iff \vec{t}) \models f_2\wedge (\vec{t}_2
\iff \vec{t})\\
{[\langle p(\vec{t}), 
f 
\rangle]}_{\equiv} 
& \mbox{otherwise} & \mbox{where }p(\vec{t}) =
msg(p(\vec{t}_1), p(\vec{t}_2)),\\
& & f = (f_1 \wedge (\vec{t}_1
\leftrightarrow \vec{t})) \lhull (f_2 \wedge (\vec{t}_2
\leftrightarrow \vec{t}))
\end{array}\right.$$
\end{proposition}

\begin{proof}
{\bf Case 1}  
$$
\begin{array}{lrcll}
& \theta(f_1) & \models & \theta(f_2) \\
\Rightarrow & (\theta(\vec{t}_1) \iff \vec{x}) \wedge \theta(f_1) &
\models & (\theta(\vec{t}_2) \iff \vec{x}) \wedge \theta(f_2) &
\mbox{by assumption}\\
\Rightarrow & (\vec{t}_1 \iff \vec{x}) \wedge \theta(f_1) & \models &
(\theta(\vec{t}_2) \iff \vec{x}) \wedge \theta(f_2) & 
\vec{t}_1 = \theta(\vec{t}_2) = \theta(\vec{t}_1)\\
\Rightarrow & (\vec{t}_1 \iff \vec{x}) \wedge f_1 & \models &
(\theta(\vec{t}_2) \iff \vec{x}) \wedge \theta(f_2) & 
var(f_1) \cap var(\vec{t}_2) = \emptyset\\
\Rightarrow & (\vec{t}_1 \iff \vec{x}) \wedge f_1 & \models &
(\vec{t}_2 \iff \vec{x}) \wedge f_2 & \models \mbox{ is transitive}\\
\Rightarrow & \restrict \vec{x}.((\vec{t}_1 \iff \vec{x}) \wedge f_1)
& \models & 
\restrict \vec{x}.((\vec{t}_2 \iff \vec{x}) \wedge f_2) &
\restrict \mbox{ is monotonic}\\
\Rightarrow & [\langle p(\vec{t}_1), f_1\rangle]_{\equiv} & \models &
[\langle p(\vec{t}_2), f_2\rangle]_{\equiv}   & \mbox{by definition}
\end{array}
$$
{\bf Case 2}
$$
\begin{array}{@{}l@{\;}r@{\;}c@{\;}ll}
& (\vec{t}_1 \iff \vec{t}) \wedge f_1 & \models &
(\vec{t}_2 \iff \vec{t}) \wedge f_2 &  \\
\Rightarrow & (\vec{t} \iff \vec{x}) \wedge (\vec{t}_1 \iff \vec{t})
\wedge f_1 & \models &  (\vec{t} \iff \vec{x}) \wedge (\vec{t}_2 \iff
\vec{t}) \wedge f_2 &  \\
\Rightarrow & \restrict \vec{x}.((\vec{t} \iff \vec{x}) \wedge (\vec{t}_1 \iff \vec{t})
\wedge f_1) & \models &  \restrict \vec{x}.((\vec{t} \iff \vec{x}) \wedge (\vec{t}_2 \iff
\vec{t}) \wedge f_2) & \restrict \mbox{ is monotonic}\\
\Rightarrow & \restrict \vec{x}.((\vec{t}_1 \iff \vec{x}) \wedge f_1)
& \models & 
\restrict \vec{x}.((\vec{t}_2 \iff \vec{x}) \wedge f_2) & \mbox{since
}\vec{x}\mbox{ are fresh}\\
\Rightarrow & [\langle p(\vec{t}_1), f_1\rangle]_{\equiv} & \models &
[\langle p(\vec{t}_2), f_2\rangle]_{\equiv}   & \mbox{by definition}
\end{array}
$$
{\bf Case 3} Immediate from lemma \ref{join-lem}.  
\end{proof}

A non-ground representation allows
chaining to be implemented efficiently using block declarations.
To check that $\land_{i = 1}^{n} y_i \lif Y_i$ entails $z \lif Z$
the variables of $Z$ are first grounded. Next, a process
is created for each clause $y_i \lif Y_i$ that suspends
until $Y_i$ is ground.
When $Y_i$ is ground, the process resumes and grounds $y_i$. If
$z$ is ground after a single pass over the clauses, then
$(\land_{i = 1}^{n} y_i \lif Y_i) \models z \lif Z$. 
Suspending and resuming a process declared by a block is constant-time
(in SICStus).
By calling
the check under negation, no problematic bindings or suspended
processes are created.

\subsection{Downward Closure}

A useful spin-off of the join algorithm in section 5.1 is
a result that shows how to calculate succinctly the
downward closure operator that arises in \mypos-based 
sharing analysis \cite{CSS99}.
Downward closure is closely related to $\lhull$ and, in fact,
$\lhull$ can be used repeatedly to compute a finite iterative sequence
that converges to $\ldown$. This is stated
in proposition~\ref{prop-down}. Finiteness follows
from bounded chain length of $\mypos_X$.

\newpage
\begin{proposition}\label{prop-down} \rm
Let $f \in \mypos_X$. Then $\ldown \!\! f = \lor_{i \geq 1} f_i$ where
$f_i \in \mypos_X$ is the increasing
chain given by: $f_1 = f$ and $f_{i+1} = f_i \lhull f_i$. 
\end{proposition}

\begin{proof}
Let $M \in \mymodel_X(\ldown \!\! f)$.
Thus there exists $M_j \in \mymodel_X(f)$
such that $M = \cup_{j = 1}^{m} M_j$. Observe
$M_1 \cap M_2, M_3 \cap M_4, \ldots \in \mymodel_X(f_2)$ and
therefore $M \in \mymodel_X(f_{\lceil \log_2(m) \rceil})$.
Since $m \leq 2^{2^n}$ where $n = |X|$ it follows
that $\ldown \!\! f \models f_{2^n}$.

Proof by induction is used for the opposite direction.
Observe that $f_1 \models \ldown \!\! f$.
Suppose $f_i \models \ldown \!\! f$. Let
$M \in \mymodel_X(f_{i+1})$. By lemma~\ref{lemma-model}
there exists $M_1, M_2 \in
\mymodel_X(f_{i})$ such that $M = M_1 \cap M_2$. By
the inductive hypothesis $M_1, M_2 \in
\mymodel_X(\ldown \!\! f)$ thus
$M \in \mymodel_X(\ldown \!\! f)$.
Hence $f_{i+1} \models \ldown \!\! f$. 

Finally, $\lor_{i \geq 1} f_i \in \mydef_X$
since $f_1 \in \mypos_X$ and
$\lhull$ is monotonic and thus
$X \in \mymodel_X(\lor_{i \geq 1} f_i)$.  
\end{proof}

The significance of this is
that it enables $\ldown$ 
to be implemented straightforwardly
with standard domain operations. This saves the implementor the
task of coding another domain operation.

\subsection{Projection}

Projection is only applied to the P component of the GEP representation
(since projection is onto the variables of the GE component).
Projection is another exponential operation.  Again, this operation
can be filtered by recognising special cases where the projection can
be calculated with lower complexity.  The projection algorithm
implemented is based on a Fourier-Motzkin style algorithm (as opposed
to a Schr\"{o}der variable elimination algorithm).
The algorithm is syntactic and
each of the variables to be projected out is
eliminated in turn. The first two steps collect clauses with the
variable to be projected out occurring in them, the third performs the
projection by syllogising and the fourth 
removes redundant clauses.  Suppose that $f = \wedge F$, where $F$ is
a set of clauses, and suppose $x$ is to be
projected out of $f$.
\begin{enumerate}
\item All those clauses with $x$ as their head are found, giving
$H = \{ x \lif X_i \mid i \in I \}$, where $I$ is a (possibly empty) index set.
\item All those clauses with $x$ in the body are found, giving
$B = \{ y \lif Y_j \mid j \in J \}$, where $J$ is a (possibly empty)
index set and $x\in Y_j$ for each $j\in J$. 
\item Let $Z_{i,j} = X_i \cup (Y_j\setminus\{x\})$.  
Then $N = \{ y \lif Z_{i,j} \mid
i \in I \land 
j \in J \land
y \not\in Z_{i,j} \}$ (syllogising).  Put $F' = ((F\setminus
H)\setminus B) \cup N$.  (Then $\exists x.f = \wedge F'$.)
\item A compact representation is
maintained by eliminating redundant clauses from $F'$ (compaction). 
\end{enumerate}
All four steps can be performed
in a single pass over $f$. A final pass over $f$ retracts
clauses such as $x \lif true$ by binding $x$ to true and also
removes clause pairs such as $y \lif z$ and $z \lif y$ by unifying $y$ and $z$.

At each pass the cost of step 4, the compaction process, is quadratic
in the size of the 
formula to be compacted (since the compaction can be reduced to a
linear number of entailment checks, each of which is linear).  The
point of compaction is to keep the representation small.  Therefore,
if the result of projecting out a variable (prior to compaction) is
smaller than the original formula, then compaction appears to be
unnecessary.  Thus, step 4 is only applied when the number of clauses
in the result of the projection is strictly greater than the number of
clauses in the original formula.  Notice also that in the filtered
case the number of syllogisms is linear in the number of occurrences
of the variable being projected out.   Table 3 details the relative
frequency with which the filtered and compaction cases are
encountered.  Observe that the vast majority of cases do not require
compaction. Finally notice that join is defined in terms of
projection, hence the 
filter for projection is inherited by join.

\begin{table}[h!]
\begin{tabular}{@{}r|@{}r|@{}r|@{}r|@{}r|@{}r|@{}r|@{}r|@{}r@{}}
file            & strips & chat\_parser & sim\_v5-2 & 
peval & aircraft & essln & chat\_80 & aqua\_c \\ \cline{1-9}
filt & 100.0 & 99.8 & 100.0 & 97.4 & 100.0 & 99.4 & 99.7 & 96.1 \\
elim &   0.0 &  0.2 &   0.0 &  2.6 &   0.0 &  0.6 &  0.3 &  3.9 
\end{tabular}
\caption{Frequency Analysis of Compaction in Projection
(induced magic)}
\end{table}

Notice that filtered algorithms break up an operation into several
components of increasing complexity.  The filtered algorithm then
suggests natural places at which to widen, i. e. 
the high complexity component is widened from above using a cheap
approximation.
This approximation might be acceptable since
the high complexity case will be called infrequently.
For example, widening might be used to improve the worst case
complexity of projection (and hence join) for non-canonical $\mydef$.

\section{Implementation of the Iteration Strategy}

Sections 3 and 4 are concerned with the 
representation of the abstract domain and the 
design and
implementation of domain operations.  The overall efficiency of an
analyser depends not only on these operations, but also on the
iteration strategy employed within the fixpoint engine.  
A fixpoint engine has to trade off the complexity of its 
data-structures against the degree of recomputation that these 
data-structures factor out.  For example, semi-na\"{\i}ve iteration \cite{ban-ram-86}
has
very simple data-structures, but entails a degree of recomputation,
whereas PLAI  \cite{her-pue-mar-stu-00} tracks dependencies
with dynamically generated graphs to dramatically reduce the amount
of recomputation.

Fixpoint engines with dependency tracking which have been applied to
logic programming analyses
include: PLAI 
\cite{her-pue-mar-stu-00}, GAIA 
\cite{LCVH94}, the CLP($\mathcal{R}$) engine \cite{kel-mar-mac-stu-yap-98} 
and GENA \cite{fec-sei-96, F97, fec-sei-99}.  An alternative to
on-the-fly dependency tracking is to use semi-na\"{\i}ve
iteration driven by a redo worklist detailing which call and answer
patterns need to be re-evaluated and (possibly) in which order.  
One instance of this is induced magic
\cite{Codish:fastmagic} under eager
evaluation \cite{Wund:lobstre95}, which factors out much of the
recomputation that arises through magic transformation. 
Other instances use knowledge of the dependencies to help order the
redo list and thereby reduce 
unnecessary computation -- this is typically done by statically
calculating SCCs \cite{gal-dew-94}, possibly recursively \cite{bou-93}, on the call graph or on the
call graph of the magic program.

The benefit of reduced recomputation is dependent upon the cost of the
abstract domain operations.  Therefore the sophistication of the
iteration strategies of engines such as PLAI and GENA is of most value
when the domain operations are complex.  The present paper has
designed its analysis so that heavyweight domain operations are
infrequently called, hence an iteration strategy employing simpler
data-structures, but possibly introducing extra computation, is
worthy of consideration.  The analyser described in \cite{how-kin-00}
used induced magic under eager evaluation.  The current analyser
builds on this work by adopting tactics inspired by PLAI, GAIA and
GENA into the induced magic framework.  Importantly these tactics
require no extra data-structures and little computational
effort. Experimental results suggest that this choice of iteration
strategy is well suited to \mydef-based groundness analysis.

\subsection{Ordered Induced Magic}

Induced magic was introduced in \cite{Codish:fastmagic}, where a
meta-interpreter for semi-na\"{\i}ve, goal-dependent, bottom-up evaluation
is presented.  The analyser described in \cite{how-kin-00}
implements a variant of this scheme using eager evaluation.  
In that paper, eager evaluation was implemented without an explicit
redo list as follows: each time a new call or answer pattern is
generated, the meta-interpreter invokes a predicate, solve,
which re-evaluates the appropriate clauses.  The re-evaluation of a
clause may in turn generate new calls to solve so that one call
may start before another finishes.  The status of these calls is
maintained on the stack, which simulates a redo list.  Henceforth,
this strategy is referred to as {\it eager induced magic}.

As noted by other authors, simple optimisations can significantly
impact on performance.  In particular, as noted in
\cite{her-pue-mar-stu-00}, evaluations resulting from new calls should
be performed before those resulting from new answers, and a call to
solve for one rule should finish before another call to solve for
another rule starts.  These optimisations cannot be integrated with
stack based eager evaluation because they rely on reordering the
calls to solve.  Hence a redo list is reintroduced in order to make these
optimisations. 

The meta-interpreter listed in Fig. 2 illustrates how a redo list can
be integrated with induced magic.  Four of the predicates are
represented as atoms in the dynamic database: {\tt redo/2}, the redo
list; {\tt fact/4}, the call and answer patterns, where propositional
formulae are represented as difference lists -- specifically, the
fourth argument is an open list with the third argument being its
tail; {\tt head\_to\_clause/2}, which represents the head and body for each
clause; {\tt atom\_to\_clause/4}, which represents the clauses with a given
atom in the body.  Before invoking {\tt oim\_solve/0}, a call to
{\tt cond\_assert/3} is required.  This has the effect of adding the
top-level call to the {\tt fact/4} database and adding the call
pattern to the {\tt redo/2} database, thereby initialising the
fixpoint calculation.  Evaluation is driven by the redo list.  If the
redo list contains call patterns, then the first (most recently
introduced) is removed and {\tt call\_solve/1} is invoked. If the redo
list contains only answer patterns, then the first is removed and
control is passed to {\tt answ\_solve/1}. The meta-interpreter
terminates (with failure) when the redo list is empty.

The predicate {\tt call\_solve/1} re-evaluates those clauses whose
heads match a new call pattern.  It first looks up a body for a clause with
a given head followed by the current call pattern for head,
then solves the body in induced magic fashion with
{\tt solve\_right/3}. If {\tt cond\_assert/3} is called with
a call (answer) pattern that does not entail the call (answer) 
pattern in {\tt fact/4}, then it succeeds, updating
{\tt fact/4} with the join of the call (answer) patterns.
In this event, the new call (answer) pattern is added to
the beginning of the {\tt redo/2} database.
The predicate {\tt answ\_solve/1} re-evaluates those clauses
containing a body atom which matches a new answer pattern.  It looks
up a clause with a body that
contains a given atom, solves the body to the left of the atom
and then to the right of the atom.
If a new call pattern is encountered in
{\tt solve\_right/4}, then the evaluation of
the clause is aborted, as the new call may
give a new answer for this body atom. In this situation, calculating
an answer for 
the head with the old body answer
will result in an answer that
needs to be re-calculated.  To ensure that the clause
is re-evaluated, an answer for the body atom
is put in the
redo list by {\tt redo\_assert/2}.       
This iteration strategy is referred to as \emph{ordered induced magic}.

\begin{figure}
\small
\begin{verbatim}
oim_solve :-
    retract(redo(call, Atom)), !, (call_solve(Atom); oim_solve).
oim_solve :-
    retract(redo(answ, Atom)), !, (answ_solve(Atom); oim_solve).

call_solve(Head) :-
    head_to_clause(Head, Body), fact(call, Head, [], Form1),
    solve_right(Body, Form1, Form2), cond_assert(answ, Head, Form2), fail. 

answ_solve(Atom) :-
    atom_to_clause(Atom, Head, Left, Right), 
    fact(call, Head, [], Form1), fact(answ, Atom, Form1, Form2),
    solve_left(Left, Form2, Form3), solve_right(Right, Form3, Form4), 
    cond_assert(answ, Head, Form4), fail.

solve_left([], Form, Form).
solve_left([Atom | Atoms], Form1, Form3) :-
    fact(answ, Atom, Form1, Form2), solve_left(Atoms, Form2, Form3).

solve_right([], Form, Form).
solve_right([Atom | Atoms], Form1, Form2) :-
    solve_right(Atom, Atoms, Form1, Form2).

solve_right(Atom, _, Form, _) :-
    cond_assert(call, Atom, Form), !, redo_assert(answ, Atom), fail.
solve_right(Atom, Atoms, Form1, Form3) :-
    fact(answ, Atom, Form1, Form2), solve_right(Atoms, Form2, Form3).
\end{verbatim}
\caption{A Meta-interpreter for Ordered Induced Magic}
\end{figure}

\subsection{SCC-based Strategies}

In order to assess the
suitability of ordered induced magic as a fixpoint strategy for
\mydef-based groundness analysis, it has been compared with a variety of
popular SCC-based methods.
The fixpoint engine can be driven either by considering 
the top-level SCCs \cite{gal-dew-94} 
or by considering the recursive nesting of SCCs, for example \cite{bou-93}.
The SCCs can be statically calculated
either on the call graph of the magicked program or on the call graph
of the original program.  

SCCs for the call graph of the magicked program (in topological order)
are calculated using
Tarjan's algorithm \cite{tar-72}.  The
fixpoint calculation then proceeds bottom-up, stabilising on the 
(call and answer)
predicates in each SCC in topological order.  If an SCC contains a
single, non-recursive, (call or answer) predicate,
then the predicate must stabilise
immediately, hence no fixpoint check is needed.  This strategy is
henceforth referred to as {\it SCC magic}.

A more sophisticated SCC-based tactic is to calculate SCCs within an
SCC, as suggested by Bourdoncle \cite{bou-93}.  
The `recursive
strategy' described by Bourdoncle recursively applies Tarjan's
algorithm to each non-trivial SCC having removed an appropriate node
(the head node)
and corresponding edges.  The fixpoint calculation proceeds
bottom-up, stabilising on the (call and answer) predicates in each
component recursively.  The fixpoint check need only be made at
the head node.  This is strategy has potential
for reaching a fixpoint in a particularly small number of
updates.  This strategy is
henceforth referred to as {\it Bourdoncle magic}.

Since both SCC magic and Bourdoncle magic work on the
call graph of the magic program, they cannot be combined with 
induced magic; the ordering of the re-evaluations conflicts.
Calculating SCCs on the call graph of the original program
may be combined with (ordered) induced magic.  The order in
which the calls are encountered is determined
by the top-down left-to-right execution of the program and  
the evaluation of a call may add new answers to the redo list.
SCCs can be used to order new answers as they are added to the redo
list.
This strategy is
henceforth referred to as {\it SCC induced magic}.
However, since calls
are re-evaluated in preference to answers, the order of answers in the
redo list is largely determined by the order of the calls. 
Consequently, SCCs should
have a negligible effect on performance.

\subsection{Dynamic Dependency Tracking}

One test of the efficacy of an iteration strategy is the number of
iterations required to reach the fixpoint.  In order to assess how
well ordered induced magic behaves, a more sophisticated iteration
strategy based on dynamic dependency tracking was implemented.   The
strategy chosen was that of WRT solver of GENA \cite{F97, fec-sei-99}
since this recent work is
particularly well described, has extensive experimental
results and conveniently fits with the redo list model.  

The WRT strategy utilises a worklist, which is effectively reordered
on-the-fly.  To quote Fecht and Seidl \cite{fec-sei-96}, ``The
worklist now is organized as a (max) priority queue where the priority
of an element [call pattern] is given by its time stamp,'' where the
time stamp records the last time the solver was called for that call pattern.
If, whilst solving for a call pattern, new call patterns are encountered,
then the bottom answer pattern is not simply returned. 
Instead the solver tries to recursively compute a
better approximation to this answer pattern.  This tactic is also
applied in PLAI and GAIA, though realised differently.

The WRT strategy of GENA gives a small number of updates, hence is an
attractive iteration strategy.  However, its implementation in a
backtrack driven meta-interpreter requires extensive use of the
dynamic database for the auxiliary data-structures.  In Prolog this is
potentially expensive \cite{HWD92}. 

\subsection{Frequency Analysis for \mydef: Reprise}

In section 3 a frequency analysis of the abstract domain operations in
\mydef-based groundness analysis was given.  It was then argued that in light
of these results certain choices about the abstract domain operations
should be made. These results are dependent on the iteration strategy
of the analyser.  In this section several different iteration
strategies have been proposed and it needs to be checked that these
give similar proportions of calls to the abstract domain operations --
that is, that the choices for the abstract domain operations remain
justified.  Table 4 gives the frequency analysis for ordered induced
magic driving non-canonical $\mydef$ and indicates that the choices of domain operation remain
valid.  Note that for the BDD analyser, each rename is
accompanied by a projection -- this is not the case for non-canonical $\mydef$,
explaining the lesser frequency of projection. This makes the non-canonical $\mydef$
representation appear even more suitable.  
Table 5 demonstrates
that projection still almost always avoids compaction. 
Similar distributions
are found with the other iteration 
strategies and for brevity these tables are omitted.

\begin{table}\label{freq-gep}
\begin{tabular}{@{}r|@{}r|@{}r|@{}r|@{}r|@{}r|@{}r|@{}r|@{}r@{}}
file            & rubik & chat\_parser & sim\_v5-2 & 
peval & aircraft & essln & chat\_80 & aqua\_c \\ \cline{1-9}
meet            & 39.3 &  40.5 & 41.5 & 44.6 &  35.4 &  48.3 &  41.0 & 43.5 \\
join            &  8.7 &   8.7 & 10.0 &  6.4 &  10.5 &   8.0 &   9.1 &  8.7 \\
join (diff)     &  1.0 &   2.0 &  0.1 &  2.6 &   0.2 &   0.7 &   1.8 &  1.3 \\
equiv           &  8.7 &   8.7 & 10.0 &  6.4 &  10.5 &   8.0 &   9.1 &  8.7 \\
proj            &  5.8 &   4.7 &  4.5 &  7.1 &   4.1 &   4.0 &   4.4 &  4.2 \\
rename          & 36.5 &  35.4 & 34.1 & 33.0 &  39.3 &  31.0 &  34.5 & 33.6\\
total           & 6646 & 11324 & 5748 & 3992 & 12550 & 11754 & 32906 & 109612
\end{tabular}
\caption{Frequency Analysis: Non-canonical $\mydef$ Analyser with Ordered
Induced Magic}
\end{table}

\begin{table}
\begin{tabular}{@{}r|@{}r|@{}r|@{}r|@{}r|@{}r|@{}r|@{}r|@{}r@{}}
file            & strips & chat\_parser & sim\_v5-2 & 
peval & aircraft & essln & chat\_80 & aqua\_c \\ \cline{1-9}
filt & 100.0 & 99.7 & 100.0 & 98.4 & 100.0 & 99.7 & 99.7 & 98.0   \\
elim & 0.0 & 0.3 & 0.0 & 1.6 & 0.0 & 0.3 & 0.3 & 2.0 
\end{tabular}
\caption{Frequency Analysis of Compaction in Projection
(Ordered Induced Magic)}
\end{table}

\section{Experimental Evaluation}

This section gives experimental results for a number of analysers with
the objective of comparing the analysis proposed in the previous
sections with existing techniques and evaluating the impact of the
various tactics utilised.
These analysers are built by selecting appropriate 
combinations of:
abstract domain, domain representation, 
iteration strategy and optimisations.
The analysers are evaluated in terms of both execution time
and the underlying behaviour (i. e. the number of updates).
All implementations are coded in SICStus Prolog 3.8.3 with the
exception of the domain operations for \mypos, which were written in C
by Schachte \cite{sch-phd}.  The analysers were run on a 296MHz Sun
UltraSPARC-II with 1GByte of RAM running Solaris 2.6.  Programs are
abstracted following the elegant (two program) scheme of
\cite{BCHP96} to guarantee correctness. Programs are normalised to definite
clauses.  Timings are the arithmetic
mean over 10 runs.
Timeouts were set at five minutes.

\subsection{Domains: Timings and Precision}

Tables 6 and 7 give timing and precision results for the domains
\myepos, \mydef \ represented in DBCF, non-canonical \mydef \ 
(denoted GEP after the representation)
and \mypos.   In these tables, file is the name of
the program 
analysed; size is the number of abstract clauses in the normalised
program; abs is the time taken to 
read, parse and normalise the input file,
producing the abstract program;
fixpoint details the analysis time for the various domains;
precision gives the total number of ground arguments in the call and answer
patterns found by each analysis (excluding those introduced by
normalising the program); \% prec. loss gives the loss of
precision of \myepos \ and \mydef \ as compared to \mypos \ -- to
emphasise where precision is lost, entries are only made when there
is a precision loss.  All the
analyses were driven by the ordered induced magic iteration strategy.

\begin{table}
\footnotesize
\begin{tabular}{@{}r|r@{\gap}r|
r@{\gap}r@{\gap}r@{\gap}r|
r@{\gap}r@{\gap}r@{\gap}r|
r@{\gap}r|}
& & & \multicolumn{4}{c|}{fixpoint}
& \multicolumn{4}{c|}{precision}
& \multicolumn{2}{c|}{\% prec. loss} \\
file & size & abs & \myepos & DBCF & GEP & \mypos & \myepos & DBCF & GEP & \mypos & \myepos & \mydef \\ \cline{1-13}
append.pl & 2 & 0.00 & 0.00 & 0.01 & 0.01 & 0.01 & 3 & 4 & 4 & 4 & 25.0 & \\
rotate.pl & 3 & 0.00 & 0.00 & 0.01 & 0.01 & 0.01 & 2 & 3 & 3 & 6 & 66.6 & 50.0 \\
mortgage.clpr & 4 & 0.00 & 0.00 & 3.31 & 0.00 & 0.04 & 6 & 6 & 6 & 6 & & \\
qsort.pl & 6 & 0.01 & 0.00 & 0.00 & 0.00 & 0.01 & 11 & 11 & 11 & 11 & & \\
rev.pl & 6 & 0.01 & 0.00 & 0.01 & 0.01 & 0.01 & 0 & 0 & 0 & 0 & & \\
queens.pl & 9 & 0.00 & 0.00 & 0.04 & 0.00 & 0.02 & 3 & 3 & 3 & 3 & & \\
zebra.pl & 9 & 0.01 & 0.00 & 0.06 & 0.01 & 0.10 & 19 & 19 & 19 & 19 & & \\
laplace.clpr & 10 & 0.01 & 0.00 & 0.08 & 0.01 & 0.01 & 0 & 0 & 0 & 0 & & \\
shape.pl & 11 & 0.00 & 0.00 & 0.04 & 0.00 & 0.03 & 6 & 6 & 6 & 6 &  & \\
parity.pl & 12 & 0.01 & 0.00 & 3.24 & 0.52 & -- & 0 & 0 & 0 & -- & -- & -- \\
treeorder.pl & 12 & 0.00 & 0.00 & 0.20 & 0.01 & 0.03 & 0 & 0 & 0 & 0 & & \\
fastcolor.pl & 13 & 0.04 & 0.00 & 0.00 & 0.01 & 0.01 & 14 & 14 & 14 & 14 & & \\
music.pl & 13 & 0.01 & 0.01 & -- & 0.02 & 0.07 & 2 & -- & 2 & 2 & & \\
serialize.pl & 13 & 0.01 & 0.00 & 0.12 & 0.00 & 0.06 & 3 & 3 & 3 & 3 & & \\
crypt\_wamcc.pl & 19 & 0.02 & 0.01 & 0.03 & 0.01 & 0.04 & 31 & 31 & 31 & 31 & & \\
option.clpr & 19 & 0.02 & 0.00 & 1.27 & 0.02 & 0.07 & 42 & 42 & 42 & 42 & & \\
circuit.clpr & 20 & 0.02 & 0.00 & 52.69 & 0.02 & 0.12 & 3 & 3 & 3 & 3 & & \\
air.clpr & 20 & 0.01 & 0.00 & 44.63 & 0.02 & 0.09 & 9 & 9 & 9 & 9 & & \\
dnf.clpr & 22 & 0.02 & 0.01 & 0.01 & 0.00 & 0.03 & 8 & 8 & 8 & 8 & & \\
dcg.pl & 23 & 0.02 & 0.00 & 0.01 & 0.00 & 0.02 & 59 & 59 & 59 & 59 & & \\
hamiltonian.pl & 23 & 0.02 & 0.00 & 0.01 & 0.00 & 0.02 & 37 & 37 & 37 & 37 &  & \\
nandc.pl & 31 & 0.03 & 0.02 & 0.03 & 0.01 & 0.05 & 34 & 37 & 37 & 37 & 8.1 & \\
semi.pl & 31 & 0.03 & 0.02 & 0.75 & 0.04 & 0.23 & 28 & 28 & 28 & 28 &  &  \\
life.pl & 32 & 0.02 & 0.00 & 0.03 & 0.01 & 0.05 & 58 & 58 & 58 & 58 &  & \\
poly10.pl & 32 & 0.03 & 0.00 & 0.02 & 0.00 & 0.04 & 45 & 45 & 45 & 45 &  & \\
meta.pl & 33 & 0.02 & 0.01 & 0.02 & 0.02 & 0.03 & 1 & 1 & 1 & 1 & & \\
rings-on-pegs.clpr & 34 & 0.02 & 0.02 & 1.20 & 0.02 & 0.11 & 11 & 11 & 11 & 11 & & \\
browse.pl & 35 & 0.02 & 0.01 & 0.04 & 0.02 & 0.04 & 41 & 41 & 41 & 41 & & \\
gabriel.pl & 38 & 0.03 & 0.02 & 0.06 & 0.02 & 0.07 & 37 & 37 & 37 & 37 & & \\
tsp.pl & 38 & 0.02 & 0.02 & 0.07 & 0.02 & 0.11 & 122 & 122 & 122 & 122 & & \\
map.pl & 41 & 0.02 & 0.02 & 0.03 & 0.01 & 0.05 & 17 & 17 & 17 & 17 &  & \\
csg.clpr & 42 & 0.04 & 0.01 & 0.00 & 0.00 & 0.02 & 8 & 8 & 8 & 8 & & \\
disj\_r.pl & 48 & 0.03 & 0.01 & 0.02 & 0.02 & 0.08 & 97 & 97 & 97 & 97 & & \\
ga.pl & 48 & 0.08 & 0.00 & 0.03 & 0.02 & 0.09 & 141 & 141 & 141 & 141 & & \\
critical.clpr & 49 & 0.03 & 0.01 & -- & 0.04 & 0.21 & 14 & -- & 14 & 14 & & \\
robot.pl & 51 & 0.04 & 0.00 & 0.01 & 0.00 & 0.03 & 41 & 41 & 41 & 41 & & \\
scc1.pl & 51 & 0.03 & 0.01 & 0.08 & 0.01 & 0.14 & 89 & 89 & 89 & 89 & & \\
ime\_v2-2-1.pl & 53 & 0.04 & 0.02 & 0.30 & 0.03 & 0.20 & 100 & 101 & 101 & 101 & 0.9 & \\
cs\_r.pl & 54 & 0.06 & 0.01 & 0.06 & 0.01 & 0.09 & 149 & 149 & 149 & 149 & & \\
tictactoe.pl & 55 & 0.05 & 0.01 & 0.08 & 0.02 & 0.09 & 60 & 60 & 60 & 60 & &  \\
flatten.pl & 56 & 0.04 & 0.02 & 0.22 & 0.04 & 0.13 & 27 & 27 & 27 & 27 &  & \\
mastermind.pl & 56 & 0.03 & 0.02 & 0.04 & 0.02 & 0.09 & 43 & 43 & 43 & 43 & & \\
dialog.pl & 61 & 0.03 & 0.01 & 0.03 & 0.02 & 0.05 & 45 & 45 & 45 & 45 &  & \\
neural.pl & 67 & 0.06 & 0.03 & 0.13 & 0.02 & 0.08 & 121 & 123 & 123 & 123 & 1.6 &  \\
bridge.clpr & 68 & 0.10 & 0.00 & 0.07 & 0.01 & 0.09 & 24 & 24 & 24 & 24 & & \\
conman.pl & 76 & 0.05 & 0.00 & 0.00 & 0.00 & 0.02 & 6 & 6 & 6 & 6 &  & \\
unify.pl & 77 & 0.05 & 0.02 & 0.19 & 0.05 & 0.38 & 70 & 70 & 70 & 70 & & \\
kalah.pl & 78 & 0.04 & 0.02 & 0.05 & 0.02 & 0.10 & 199 & 199 & 199 & 199 & & \\
nbody.pl & 85 & 0.07 & 0.03 & 0.08 & 0.04 & 0.19 & 113 & 113 & 113 & 113 & & \\
peep.pl & 85 & 0.11 & 0.02 & 0.13 & 0.03 & 0.14 & 10 & 10 & 10 & 10 &  & \\
sdda.pl & 89 & 0.05 & 0.02 & 0.13 & 0.04 & 0.12 & 17 & 17 & 17 & 17 & &  \\
bryant.pl & 94 & 0.07 & 0.06 & 0.23 & 0.14 & 0.76 & 99 & 99 & 99 & 99 & & \\
boyer.pl & 95 & 0.07 & 0.02 & 0.08 & 0.05 & 0.08 & 3 & 3 & 3 & 3 & & \\
read.pl & 101 & 0.09 & 0.03 & 0.15 & 0.05 & 0.20 & 99 & 99 & 99 & 99 &  & \\
qplan.pl & 108 & 0.09 & 0.02 & 0.07 & 0.02 & 0.16 & 216 & 216 & 216 & 216 & & \\
trs.pl & 108 & 0.14 & 0.06 & -- & 0.09 & 2.46 & 13 & -- & 13 & 13 &  &  \\
press.pl & 109 & 0.08 & 0.07 & 0.40 & 0.10 & 0.36 & 52 & 53 & 53 & 53 & 1.8 &  \\
reducer.pl & 113 & 0.07 & 0.05 & 3.47 & 0.04 & 0.30 & 41 & 41 & 41 & 41 &  &  \\
parser\_dcg.pl & 122 & 0.09 & 0.04 & 2.27 & 0.08 & 0.24 & 28 & 43 & 43 & 43 & 34.8 &  \\
simple\_analyzer.pl & 140 & 0.11 & 0.05 & 0.28 & 0.10 & 0.58 & 89 & 89 & 89 & 89 & &  \\
\end{tabular}
\caption{Groundness Results: Smaller Programs}
\end{table}

\begin{table}
\footnotesize
\begin{tabular}{@{}r|r@{\gap}r|
r@{\gap}r@{\gap}r@{\gap}r|
r@{\gap}r@{\gap}r@{\gap}r|
r@{\gap}r|}
& & & \multicolumn{4}{c|}{fixpoint}
& \multicolumn{4}{c|}{precision}
& \multicolumn{2}{c|}{\% prec. loss} \\
file & size & abs & \myepos & DBCF & GEP & \mypos & \myepos & DBCF & GEP & \mypos & \myepos & \mydef \\ \cline{1-13}
dbqas.pl & 143 & 0.09 & 0.02 & 0.54 & 0.03 & 0.09 & 18 & 18 & 18 & 18 & & \\
ann.pl & 146 & 0.10 & 0.05 & 0.77 & 0.09 & 0.32 & 71 & 71 & 71 & 71 & & \\
asm.pl & 160 & 0.17 & 0.04 & 0.08 & 0.04 & 0.17 & 90 & 90 & 90 & 90 & & \\
nand.pl & 179 & 0.14 & 0.04 & 0.19 & 0.05 & 0.37 & 402 & 402 & 402 & 402 & &  \\
lnprolog.pl & 220 & 0.10 & 0.07 & 0.16 & 0.07 & 0.21 & 110 & 143 & 143 & 143 & 23.0 & \\
ili.pl & 221 & 0.15 & 0.07 & 1.29 & 0.17 & 0.36 & 4 & 4 & 4 & 4 &  & \\
strips.pl & 240 & 0.22 & 0.02 & 0.04 & 0.03 & 0.14 & 142 & 142 & 142 & 142 & & \\
sim.pl & 244 & 0.20 & 0.08 & 1.69 & 0.18 & 1.38 & 100 & 100 & 100 & 100 & & \\
rubik.pl & 255 & 0.20 & 0.12 & -- & 0.16 & 0.46 & 158 & -- & 158 & 158 & &  \\
chat\_parser.pl & 281 & 0.34 & 0.09 & 0.47 & 0.24 & 1.16 & 504 & 505 & 505 & 505 & 0.1 &  \\
sim\_v5-2.pl & 288 & 0.23 & 0.05 & 0.15 & 0.07 & 0.32 & 455 & 455 & 455 & 457 & 0.4 & 0.4 \\
peval.pl & 332 & 0.17 & 0.05 & 0.23 & 0.18 & 0.39 & 27 & 27 & 27 & 27 &  &  \\
aircraft.pl & 395 & 0.55 & 0.11 & 0.21 & 0.14 & 0.55 & 687 & 687 & 687 & 687 &  &  \\
essln.pl & 595 & 0.48 & 0.12 & 2.70 & 0.19 & 0.93 & 158 & 162 & 162 & 162 & 2.4 & \\
chat\_80.pl & 883 & 1.53 & 0.38 & 8.17 & 0.76 & 4.53 & 852 & 855 & 855 & 855 & 0.3 & \\
aqua\_c.pl & 3928 & 3.47 & 1.70 & -- & 4.26 & 144.62 & 1222 & -- & 1285 & 1285 & 4.9 &  \\
\end{tabular}
\caption{Groundness Results: Larger Programs}
\end{table}

First consider precision.  As is well known, in practice, for
goal-dependent groundness analysis, the
precision of \mydef \ is very close to that of \mypos.  In the benchmark suite
used here, \mydef \ loses ground arguments in only two programs:
rotate.pl, which loses three arguments, and
sim\_v5-2.pl, where two arguments are lost.  \myepos \ loses
precision in several programs, but still performs reasonably well.
(Goal-independent analysis precision comparisons for $\myepos$ and
$\mydef$ are given in \cite{HACK00} and \cite{gen-cod-01}. These
show that $\myepos$ loses significant precision, whereas
$\mydef$ gives precision close to that of $\mypos$.)

The non-canonical \mydef \ analyser appears to be fast and
scalable -- taking 
more than a second to analyse only the largest benchmark program.
This analyser does not employ widening (however, 
incorporating a widening 
would guarantee robustness of the analyser, even for pathological
programs \cite{gen-how-cod-01}).
Notice that the analysis times for all the programs is close to the
abstraction time -- this suggests that a large speed up in the
analysis time needs to be coupled with a commensurate speedup in the
abstracter.

The non-canonical \mydef \ analysis times are comparable to those for \myepos \ for
smaller programs, with \myepos \ outperforming non-canonical \mydef \ on some of the
larger benchmarks.  This is unsurprising given the much better theoretical
behaviour of \myepos, indeed it is much in the
favour of non-canonical \mydef \ that it is competitive with \myepos.
The DBCF analyser suffers from the problems discussed in section 4.
The cost in meet of maintaining the canonical form often becomes
significant.  In cases (such as in music.pl) where the number of
variables, the number of body atoms and the size of the representation
are all large, the exponential nature of reducing to canonical form leads
to a massive blowup in analysis time. Hence the DBCF analyser fails to
produce a result for several examples and gives poor scalability.
Also, the analysis appears to lack robustness -- the sensitivity of
the meet to the form of the program clauses leads to widely varying
results.
\mypos \ performs well on most programs, but is still consistently
several times 
slower than non-canonical \mydef.  \mypos \ performs
particularly poorly on parity.pl (a program designed to be
problematic for BDD-based \mypos \ analysers) and aqua\_c.pl.  Again, since the
\mypos \ analyser uses BDDs (essentially a canonical form)
there is a cost in maintaining the representation.  This can lead to a
lack of robustness.  
It should be pointed out that the \mypos \
analyser is not state of the art and that one using the GER representation
\cite{BS98} would probably give improved results.  Of course, widening could be
used to give improved times for \mypos, but at the cost of precision.

\subsection{Iteration Strategy: Timings and Updates}

Table 8 gives timing results for non-canonical \mydef \ analysis when driven by  various
iteration strategies.  The column headers are abbreviations as
follows: ord stand for ordered induced magic; eim stands for eager
induced magic; bom stands for Bourdoncle magic; scm stands for SCC
magic; scc stands for SCC induced magic; dyd
stands for dynamic dependency.  The timings are split into two
sections.  The overhead time is the preprocessing overhead incurred in 
calculating the SCCs required to drive the analyses.  For bom and scm, 
SCCs are calculated on the call and answer
graph of the magic program.  For scc, SCCs are calculated on the call
graph  of the original program.
The strategies ord, eim and dyd do not require any preprocessing, hence
have no overhead.  The strategy times are the 
times for analysing each program (that is, the time taken for the
fixpoint calculation, not including the preprocessing overhead).
Table 9 gives a second measure of the cost of each iteration strategy;
this time in terms of the number of updates (writes to
database/extension table) required to reach the
fixpoint.

\begin{table}
\begin{tabular}{r|rrr|rrrrrr|}
& \multicolumn{3}{c|}{overhead}
& \multicolumn{6}{c|}{strategy} \\
file & bom & scm & scc & ord & eim & bom & scm & scc & dyd \\
\cline{1-10}
dbqas.pl & 0.02 & 0.02 & 0.01 & 0.03 & 0.03 & 0.03 & 0.06 & 0.03 & 0.07 \\
ann.pl & 0.05 & 0.04 & 0.01 & 0.09 & 0.14 & 0.18 & 0.22 & 0.09 & 0.19 \\
asm.pl & 0.06 & 0.06 & 0.02 & 0.04 & 0.08 & 0.09 & 0.13 & 0.05 & 0.15 \\
nand.pl & 0.10 & 0.08 & 0.02 & 0.05 & 0.06 & 0.21 & 0.13 & 0.05 & 0.17 \\
lnprolog.pl & 0.07 & 0.06 & 0.03 & 0.07 & 0.10 & 0.23 & 0.19 & 0.07 & 0.22 \\
ili.pl & 0.06 & 0.04 & 0.02 & 0.17 & 0.29 & 0.73 & 0.38 & 0.16 & 0.68 \\
strips.pl & 0.10 & 0.08 & 0.03 & 0.03 & 0.01 & 0.10 & 0.06 & 0.03 & 0.07 \\
sim.pl & 0.10 & 0.07 & 0.02 & 0.18 & 0.35 & 0.38 & 0.29 & 0.19 & 0.37 \\
rubik.pl & 0.29 & 0.15 & 0.04 & 0.16 & 0.19 & 1.12 & 0.33 & 0.15 & 0.34 \\
chat\_parser.pl & 0.19 & 0.08 & 0.05 & 0.24 & 0.44 & 2.31 & 0.67 & 0.24 & 1.89 \\
sim\_v5-2.pl & 0.25 & 0.12 & 0.04 & 0.07 & 0.07 & 0.57 & 0.18 & 0.07 & 0.22 \\
peval.pl & 0.06 & 0.06 & 0.04 & 0.18 & 0.30 & 0.31 & 0.29 & 0.17 & 0.38 \\
aircraft.pl & 0.73 & 0.26 & 0.14 & 0.14 & 0.23 & 1.13 & 0.53 & 0.13 & 0.44 \\
essln.pl & 0.40 & 0.19 & 0.11 & 0.19 & 0.27 & 1.58 & 0.61 & 0.18 & 0.46 \\
chat\_80.pl & 0.96 & 0.34 & 0.15 & 0.76 & 1.36 & 21.22 & 2.59 & 0.73 & 3.30 \\
aqua\_c.pl & 17.91 & 1.59 & 0.84 & 4.26 & 10.69 & 454.22 & 20.52 & 4.30 & 15.74 \\
\end{tabular}
\caption{Timing Results for Iteration Strategies}
\end{table}

\begin{table}
\footnotesize
\begin{tabular}{@{\hspace{-100pt}}l@{\hspace{-190pt}}l}
\begin{tabular}{@{}r@{\gapb}|@{\gapa}r@{\gapa}r@{\gapa}r@{\gapa}r@{\gapa}r@{\gapa}r@{\gapb}|}
& \multicolumn{6}{c|}{strategy} \\
file 
& ord & eim & bom & scm & scc & dyd \\ \cline{1-7}
append & 3 & 3 & 3 & 3 & 3 & 3 \\
rotate & 7 & 7 & 7 & 7 & 7 & 6 \\
mortgage & 6 & 6 & 6 & 6 & 6 & 6 \\
qsort & 8 & 7 & 8 & 8 & 8 & 7 \\
rev & 11 & 11 & 11 & 11 & 11 & 11 \\
queens & 12 & 12 & 12 & 12 & 12 & 12 \\
zebra & 12 & 12 & 12 & 12 & 12 & 12 \\
laplace & 12 & 12 & 12 & 12 & 12 & 12 \\
shape & 12 & 10 & 10 & 10 & 12 & 10 \\
parity & 38 & 47 & 38 & 38 & 38 & 37 \\
treeorder & 17 & 18 & 17 & 18 & 17 & 14 \\
fastcolor & 18 & 19 & 18 & 18 & 18 & 18 \\
music & 13 & 13 & 13 & 12 & 13 & 13 \\
serialize & 16 & 18 & 16 & 16 & 16 & 10 \\
crypt\_wamcc & 23 & 23 & 23 & 23 & 23 & 23 \\
option & 30 & 35 & 30 & 30 & 30 & 29 \\
circuit & 32 & 31 & 30 & 34 & 32 & 29 \\
air & 32 & 35 & 32 & 36 & 32 & 29 \\
dnf & 8 & 8 & 8 & 8 & 8 & 8 \\
dcg & 31 & 30 & 30 & 30 & 31 & 30 \\
hamiltonian & 28 & 28 & 28 & 28 & 28 & 28 \\
nandc & 49 & 51 & 44 & 51 & 49 & 49 \\
semi & 53 & 51 & 51 & 54 & 53 & 48 \\
life & 30 & 30 & 30 & 31 & 30 & 30 \\
poly10 & 24 & 24 & 24 & 24 & 24 & 24 \\
meta & 46 & 29 & 40 & 40 & 46 & 40 \\
rings-on-pegs & 37 & 37 & 37 & 37 & 37 & 37 \\
browse & 43 & 43 & 43 & 43 & 43 & 43 \\
gabriel & 48 & 48 & 48 & 50 & 48 & 47 \\
tsp & 66 & 66 & 65 & 73 & 66 & 65 \\
map & 68 & 68 & 68 & 68 & 68 & 68 \\
csg & 12 & 12 & 12 & 12 & 12 & 12 \\
disj\_r & 58 & 58 & 58 & 58 & 58 & 58 \\
ga & 60 & 60 & 59 & 60 & 60 & 59 \\
critical & 42 & 39 & 44 & 44 & 42 & 36 \\
robot & 28 & 28 & 28 & 28 & 28 & 28 \\
scc1 & 51 & 50 & 50 & 50 & 51 & 50 \\
ime\_v2-2-1 & 77 & 74 & 72 & 77 & 77 & 70 \\
\end{tabular}

&

\begin{tabular}{@{}r@{\gapb}|@{\gap}r@{\gap}r@{\gap}r@{\gap}r@{\gap}r@{\gap}r@{\gapb}|}
& \multicolumn{6}{c|}{strategy} \\
file 
& ord & eim & bom & scm & scc & dyd \\ \cline{1-7}
cs\_r & 66 & 66 & 66 & 66 & 66 & 66 \\
tictactoe & 60 & 56 & 56 & 57 & 60 & 55 \\
flatten & 81 & 95 & 80 & 107 & 81 & 71 \\
mastermind & 86 & 84 & 82 & 85 & 86 & 82 \\
dialog & 82 & 95 & 79 & 82 & 82 & 77 \\
neural & 83 & 78 & 78 & 102 & 83 & 78 \\
bridge & 13 & 13 & 13 & 13 & 13 & 13 \\
conman & 14 & 14 & 14 & 14 & 14 & 14 \\
unify & 92 & 114 & 92 & 97 & 92 & 83 \\
kalah & 91 & 93 & 92 & 93 & 91 & 92 \\
nbody & 125 & 173 & 124 & 162 & 125 & 122 \\
peep & 61 & 61 & 62 & 61 & 61 & 58 \\
sdda & 91 & 105 & 96 & 100 & 93 & 94 \\
bryant & 202 & 210 & 189 & 161 & 202 & 214 \\
boyer & 99 & 107 & 102 & 101 & 99 & 105 \\
read & 119 & 127 & 90 & 114 & 119 & 91 \\
qplan & 95 & 95 & 95 & 94 & 95 & 93 \\
trs & 86 & 92 & 88 & 96 & 88 & 69 \\
press & 224 & 222 & 221 & 217 & 224 & 241 \\
reducer & 118 & 173 & 173 & 158 & 118 & 163 \\
parser\_dcg & 170 & 170 & 157 & 168 & 169 & 160 \\
simple\_analyzer & 200 & 242 & 200 & 321 & 201 & 189 \\
dbqas & 105 & 105 & 94 & 109 & 105 & 98 \\
ann & 207 & 233 & 229 & 281 & 207 & 192 \\
asm & 169 & 237 & 174 & 217 & 169 & 181 \\
nand & 188 & 188 & 186 & 187 & 188 & 186 \\
lnprolog & 253 & 300 & 279 & 281 & 253 & 264 \\
ili & 209 & 318 & 318 & 330 & 209 & 312 \\
strips & 108 & 101 & 111 & 106 & 108 & 99 \\
sim & 280 & 310 & 269 & 277 & 281 & 266 \\
rubik & 372 & 369 & 375 & 383 & 372 & 373 \\
chat\_parser & 445 & 682 & 659 & 652 & 445 & 621 \\
sim\_v5-2 & 256 & 256 & 254 & 254 & 256 & 256 \\
peval & 280 & 331 & 312 & 309 & 281 & 285 \\
aircraft & 506 & 506 & 506 & 506 & 506 & 506 \\
essln & 485 & 547 & 473 & 516 & 485 & 450 \\
chat\_80 & 1322 & 1657 & 1494 & 1579 & 1323 & 1454 \\
aqua\_c & 4751 & 5779 & 5667 & 6106 & 4842 & 4611 \\
\end{tabular}
\end{tabular}
\caption{Number of Updates for Iteration Strategies}
\end{table}

One important measure of the success of an iteration
strategy is the number of updates required in the analysis.  This
impacts directly on the number of calls to abstract operations
and hence the amount of work (speed) of the analysis.  Table 9
indicates that ord, scc and dyd give the best behaviour over a large number
of programs.  However, all of the other strategies give the best
result for some programs, indicating that each has its merits.
Observe that, as predicted in section 5, ord and scc give very similar
results. 

In measuring performance of a particular analysis, the 
overall time taken is also of importance.  Table 8 indicates that the methods
based on SCCs in the call graph of the magic program have
problems. Firstly, they require SCCs to
be calculated -- the cost of this (in particular for 
Bourdoncle magic) is significant.
Secondly, the fixpoint times for bom and
scm are much greater than would be expected from the results in Table 9.  
This is partly because the bom and scm strategies
cannot be integrated with induced magic, which 
impacts heavily on speed.  The bom strategy also has a third drawback
-- the proportion of re-evaluations not resulting in an update
rises dramatically for larger programs.  Larger programs often give
rise to deeply nested SCCs.
Suppose an SCC, say $A$, nests a subSCC, say $B$.  In detecting the
stability of $A$, the stability of the head of $B$ needs to be
established.  This in turn requires a single pass over $B$.
If $n$ passes over $A$ are required to reach stability, then $n$ passes
over $B$ are also needed (even if $B$ is already stable).
Extrapolating, the number of times an SCC is 
passed over is determined by the sum of the number of passes over each SCC
containing it.  If the SCC is deeply nested and large this involves a large
number of re-evaluations producing no updates. As the scm strategy does
not involve nested SCCs, this problem does not arise.
It appears that Bourdoncle's recursive strategy is not well suited for
driving  
groundness analyses of logic programs.  
Table 8 also indicates that
whilst SCCs on the call graph give comparable analysis times to
ordered induced magic, they too come with an overhead of
precomputation.  
Sophisticated dynamic dependency graphs do not pay for
themselves in a groundness analysis involving lightweight domain operations,
as reflected by the timings for dyd.
However, they are more amenable to optimisation than ordered induced
magic (which is itself essentially an optimisation of induced magic)
and in an analysis where the cost of the
abstract operations is higher it is to be expected that this strategy
would be more effective.  
Also, by using a different programming
paradigm, the dynamic changes to the dependency graph could be made
more efficiently (for example, \cite{fec-sei-99} use SML).

\subsection{Chain Length}

Table 10 gives further details of the number of updates required in 
program analysis with non-canonical \mydef.
This table gives the distribution of the number of updates
required to reach the fixpoint for the various program
predicates. Results are given for ord and dyd as it is clear from
Table 9 that these are the most competitive strategies.  Each column
gives the number of predicates requiring that number of updates.
Entries beyond the maximum number of
updates have been left blank to highlight the maximum chain length.

\begin{table}
\begin{tabular}{r|r@{\gap}r@{\gap}r@{\gap}r@{\gap}
r@{\gap}r@{\gap}r@{\gap}r@{\gap}r|r@{\gap}r@{\gap}r@{\gap}r@{\gap}r@{\gap}r@{\gap}r@{\gap}r|}
& \multicolumn{9}{c|}{ord}
& \multicolumn{8}{c|}{dyd} \\
file & 1 & 2 & 3 & 4 & 5 & 6 & 7 & 8 & 9 & 1 & 2 & 3 & 4 & 5 & 6 & 7 & 8 \\
\cline{1-18}
dbqas.pl & 55 & 21 & 1 & 0 & 1 & & & &  & 60 & 16 & 2 & & & & &   \\
ann.pl & 88 & 39 & 11 & 2 & & & & &  & 100 & 32 & 4 & 4 & & & &   \\
asm.pl & 140 & 13 & 1 & & & & & & & 130 & 21 & 3 & & & & &  \\
nand.pl & 172 & 5 & 2 & & & & & & & 173 & 5 & 1 & & & & &  \\
lnprolog.pl & 168 & 32 & 7 & & & & & &  & 155 & 47 & 5 & & & & &  \\
ili.pl & 89 & 24 & 20 & 3 & & & & &  & 41 & 54 & 23 & 5 & 5 & 7 & 1 &  \\
strips.pl & 82 & 10 & 2 & & & & & & & 89 & 5 & & & & & &  \\
sim.pl & 144 & 43 & 12 & 1 & 2 & & & & & 152 & 38 & 10 & 2 & & & &  \\
rubik.pl & 264 & 54 & & & & & & & & 264 & 53 & 1 & & & & &  \\
chat\_parser.pl & 207 & 78 & 14 & 7 & 1 & 0 & 1 & & & 101 & 144 & 37 & 14 & 7 & 5 & &  \\
sim\_v5-2.pl & 248 & 4 & & & & & & & & 248 & 4 & & & & & &  \\
peval.pl & 114 & 45 & 13 & 4 & 3 & 1 & & & & 111 & 52 & 9 & 3 & 2 & 1 & 1 & 1 \\
aircraft.pl & 468 & 19 & & & & & & & & 468 & 19 & & & & & &  \\
essln.pl & 321 & 59 & 9 & 2 & 1 & 1 & & & & 341 & 48 & 3 & 1 & & & &  \\
chat\_80.pl & 537 & 224 & 70 & 22 & 4 & 2 & 1 & & & 466 & 261 & 93 & 22 & 10 & 7 & 1 &  \\
aqua\_c & 2135 & 742 & 205 & 64 & 28 & 12 & 2 & 1 & 3 & 2170 & 781 & 151 & 48 & 26 & 11 & 2 & 3 \\
\end{tabular}
\caption{Chain Length Distributions}
\end{table}

Chain length gives a good indication of the robustness of the
iteration strategies.  Whilst it is always possible to construct
programs exhibiting worst case behaviour \cite{cod-00,gen-how-cod-01}, 
Table 10 shows that for
both ord and  dyd, very
few chains are longer than 4 and that at worst chains have length 9.
It also again indicates that different strategies can give
significantly different behaviour for the analysis.

\subsection{Optimisations}

A number of optimisations have been discussed in this paper.
Table 11 details the effect of these, singly and in combination.  The
five optimisations considered have each been abbreviated by a single
letter:  e denotes filtering by entailment checking; g denotes the use
of a GEP factorisation; p denotes filtering projection; r denotes the
use of redundancy removal; t denotes the maintenance of a true factorisation.
The column headers describe which optimisations have
been switched on; for example, gpr denotes the situation where the
analysis uses a GEP factorisation, where projection is filtered and
where redundancy removal is used, but the factorisation is not
true and the entailment checking filter for join is not applied.  
Note that the switch for the entailment checking does
not entirely turn off the entailment check filter for join, as the \mydef \
analysers enforce termination using the same 
entailment check which filters join.
In Proposition 2, the filtering
of join has three cases; the entailment check switch turns the first
(most lightweight) case on and off.  The default for the non-canonical
\mydef \ analyser
which has been used for other timings in this paper is egpr, since
this gives the best result for most programs.

\begin{table}
\begin{tabular}{r|rrrrrrrrrr|}
& \multicolumn{10}{c|}{switches} \\
file & egpr & egprt & egp & epr & gpr & egr & pr & er & gr & r \\
\cline{1-11}
dbqas.pl & 
0.02 & 0.02 & 0.02 & 0.03 & 0.03 & 0.02 & 0.03 & 0.03 & 0.03 & 0.03 \\
ann.pl &
0.09 & 0.09 & 0.09 & 0.10 & 0.10 & 0.09 & 0.11 & 0.10 & 0.10 & 0.11 \\
asm.pl &
0.04 & 0.04 & 0.04 & 0.05 & 0.04 & 0.04 & 0.05 & 0.05 & 0.04 & 0.06 \\
nand.pl & 
0.05 & 0.05 & 0.05 & 0.07 & 0.06 & 0.05 & 0.09 & 0.07 & 0.06 & 0.09 \\
lnprolog.pl & 
0.06 & 0.07 & 0.07 & 0.08 & 0.07 & 0.07 & 0.09 & 0.08 & 0.07 & 0.10 \\
ili.pl & 
0.16 & 0.16 & 0.16 & 0.17 & 0.16 & 0.16 & 0.18 & 0.17 & 0.18 & 0.19 \\
strips.pl & 
0.02 & 0.02 & 0.02 & 0.03 & 0.03 & 0.02 & 0.04 & 0.03 & 0.03 & 0.03 \\
sim.pl &
0.18 & 0.18 & 0.18 & 0.20 & 0.20 & 0.20 & 0.23 & 0.23 & 0.22 & 0.25 \\
rubik.pl & 
0.15 & 0.16 & 0.16 & 0.16 & 0.17 & 0.19 & 0.18 & 0.20 & 0.19 & 0.21 \\
chat\_parser.pl & 
0.24 & 0.24 & 0.24 & 0.29 & 0.27 & 0.25 & 0.33 & 0.30 & 0.28 & 0.34 \\
sim\_v5-2.pl & 
0.06 & 0.07 & 0.06 & 0.08 & 0.07 & 0.07 & 0.10 & 0.08 & 0.07 & 0.10 \\
peval.pl & 
0.17 & 0.17 & 0.17 & 0.18 & 0.17 & 0.18 & 0.19 & 0.19 & 0.18 & 0.20 \\
aircraft.pl & 
0.14 & 0.14 & 0.14 & 0.17 & 0.16 & 0.14 & 0.21 & 0.17 & 0.16 & 0.21 \\
essln.pl & 
0.18 & 0.19 & 0.19 & 0.22 & 0.20 & 0.19 & 0.24 & 0.22 & 0.21 & 0.25 \\
chat\_80.pl & 
0.73 & 0.74 & 0.73 & 0.84 & 0.81 & 0.76 & 0.96 & 0.89 & 0.84 & 0.99 \\
aqua\_c.pl & 
4.25 & 4.20 & 4.28 & 4.74 & 4.73 & 4.81 & 5.34 & 5.36 & 5.29 & 5.99 \\
\end{tabular}
\caption{Timing Results for Combinations of Optimisations}
\end{table}

The first three columns of Table 11 all give very similar times,
indicating that true factorisation and redundancy removal have little
effect on analysis times, essentially paying for themselves.  The next
three columns give times for the situation with one of e, g, p
switches off (relative to the default case).  It is clear that turning
off any of these optimisation gives a slow down of, perhaps, 10\%.
The next three columns give results for switching off optimisations in
pairs.  Again there is a clear slowdown from the previous three
results (although notice that the epr and gr results are very
similar), a slowdown of 15-20\% from the default case. Finally, the
last column shows that switching off all the optimisations results in
a slowdown of approximately 25\% in most programs.

One conclusion to be drawn from Table 11 is that the non-canonical \mydef\ analysis is
extremely robust.  By turning off all the optimisations for both the size
of representation and the efficiency of the abstract operations, the
analysis is still fast.  It is expected that the effect of
turning off these optimisations would be bigger when using a less
effective iteration strategy or a less suitable (orthogonal) representation.

\section{Related Work}

{Van Hentenryck} {\it et al.} \cite{CLCVH95} is an early
work which laid a foundation for BDD-based
$\mypos$ analysis.
Corsini {\it et al.} \cite{CMRLC93} describe how variants of $\mypos$ can be
implemented using Toupie, a constraint language based on the
$\mu$-calculus. If this analyser was extended with, say, magic
sets, it might lead to a very respectable goal-dependent analysis.
More recently, Bagnara and Schachte \cite{BS98} have
developed the idea \cite{B96} that a factorised
implementation of ROBDDs which keeps definite information
separately from dependency information is more efficient than
keeping the two together. This hybrid representation can
significantly decrease
the size of an ROBDD and thus is a useful implementation tactic. 

Heaton {\it et al.} \cite{HACK00} propose \myepos, a sub-domain of $\mydef$,
that can only propagate
dependencies of the form $(x_1 \liff x_2) \land x_3$ across
procedure boundaries.  
This information is precisely that contained in one of the fields of
the GEP factorised domain.
The main finding of 
\cite{HACK00} is that this sub-domain performs reasonably  well 
for goal-dependent analysis.

Armstrong {\it et al.} \cite{AMSS98} study a number of
different representations of Boolean functions
for both $\mydef$ and $\mypos$. An empirical evaluation on
15 programs suggests that specialising Dual Blake Canonical Form (DBCF)
for $\mydef$ leads to the fastest analysis overall. 
Armstrong {\it et al.} \cite{AMSS98} also perform interesting
precision experiments. $\mydef$ and $\mypos$ are compared, however, in a 
bottom-up framework that is based
on condensing and is therefore biased towards $\mypos$.
The authors point out that a top-down analyser would
improve the precision of $\mydef$ relative to $\mypos$.

Garc\'{\i}a de la Banda {\it et al.} \cite{BHBDJS96} describe
a Prolog implementation of
$\mydef$ that is also based on an orthogonal DBCF representation
(though this is not explicitly stated) and show that it
is viable for some medium sized benchmarks.
Fecht and Seidl \cite{F97, fec-sei-99} describe another groundness
analyser  for \mypos \ that is
not coded in C. They adopt SML as a coding medium in order to build
an analyser that is declarative and easy to maintain. 
Their analyser employs a widening.

Codish and Demoen \cite{CD95} describe a non-ground model based
implementation technique
for $Pos$ that would encode 
$x_1 \liff (x_2 \land x_3)$ as three
tuples 
$\langle true, true, true \rangle$,
$\langle false, \_, false \rangle$,
$\langle false, false, \_ \rangle$. 
King {\it et al.} show how, for \mydef,  meet, join and projection
can be implemented with quadratic operations based on a
$\myshare$ quotient \cite{KSH99}. $\mydef$ functions are
essentially represented as a set of models and widening is
thus required to keep the size of the
representation manageable. 
Ideally, however, it would be better
to avoid widening by, say,
using a more compact representation.

Most recently, Genaim and Codish \cite{gen-cod-01} propose a dual
representation for \mydef.  For function $f$, the 
models of $coneg(f)$ are named and $f$ is represented by a tuple recording for
each variable of $f$ which of these models the variable is in.  For
example, the models of $coneg(x\imp 
y)$ are  $\{\{x, y\}, \{x\}, \emptyset\}$.  Naming the three models $a$,
$b$, $c$ respectively, $f$ is represented by $\langle ab, a \rangle$.
This representation cleverly allows the well known ACI1 unification
theory to be used for the domain operations.  \cite{gen-cod-01} report
promising experimental results, but still need a widening to analyse
the aqua\_c benchmark.

\section{Conclusion}

By considering the way in which goal-dependent groundness analyses
proceed, an intelligent choice can be made as to how to represent the
abstract domain and how the cost of the domain operations should be
balanced.  Analysing the relative frequencies of the domain operations leads to
a representation which is compact, and where the most commonly called domain
operations are the most lightweight.  Filters for the more expensive
domain operations are described which allow these operations to be
calculated by inexpensive special cases.  
Ways in which
a non-ground representation for Boolean functions may exploit the
language features of Prolog to obtain an efficient
implementation are described.  
The iteration strategy for driving an analysis is
also extremely important.  Several strategies are discussed 
and compared.
It is concluded that for groundness analysis the fastest
implementation 
uses a simple strategy avoiding 
precomputation and sophisticated data-structures.
An implementor might find some or all of the issues discussed and
ideas raised in this
paper useful in designing a program analysis and in implementing it in
Prolog. 

The end product of this work is a highly principled
goal-dependent groundness analyser  
combining  the techniques described.  It is written in Prolog and
is small and easily maintained.  The analyser 
is a robust, fast, precise and scalable
and does not require widening for the largest program in the benchmark
suite.  Experimental results show that the speed of the fixpoint
calculation is very close to that of reading, parsing and normalising
the input file.  Results also
suggest that the performance of the analyser
compares well with other groundness analysers, including
BDD-based analysers written in C.

\subsubsection*{Acknowledgements}

We thank Roberto Bagnara, Fran\c{c}ois Bourdoncle, \mbox{Mike Codish},
\mbox{Roy Dyckhoff}, 
John Gallagher, Samir Genaim and Pat Hill for useful discussions. We
would also like to thank Peter Schachte for help with his BDD analyser.
This work was funded partly by EPSRC Grant GR/MO8769.


\end{document}